\DeclareRobustCommand{\VAN}[3]{#2}
\let\VANthebibliography\thebibliography
\def\thebibliography{\DeclareRobustCommand{\VAN}[3]{##3}\VANthebibliography}

\documentclass[fleqn]{mnras}

\usepackage{graphicx}	
\usepackage{amsmath}	
\usepackage{amssymb}

\graphicspath{{plots/}}
\usepackage[linesnumbered,ruled,vlined]{algorithm2e}
\usepackage{bm}		
\usepackage{pdflscape}	
\usepackage{ulem}
\let\bibhang\relax              
\usepackage{natbib}
\usepackage{makecell}
\usepackage{array}
\usepackage{booktabs}
\usepackage[utf8]{inputenc}

\newcommand{\hGpc}{\,h^{-1}~{\rm Gpc}}
\newcommand{\Gpc}{\,{\rm Gpc}}
\newcommand{\Mpc}{\,{\rm Mpc}}
\newcommand{\hMpc}{{\ifmmode{h^{-1}{\rm Mpc}}\else{$h^{-1}$Mpc}\fi}}
\newcommand{\hkpc}{{\ifmmode{h^{-1}{\rm kpc}}\else{$h^{-1}$kpc}\fi}}
\newcommand{\hMsun}{{\ifmmode{\,h^{-1}{\rm {M_{\odot}}}}\else{$h^{-1}{\rm{M_{\odot}}}$}\fi}}
\newcommand{\Msun}{\,\rm {M_{\odot}}}
\newcommand{\Mhalo}{{\ifmmode{\,M_{\rm halo}}\else{$M_{\rm halo}$}\fi}}
\newcommand{\ltsima}{$\; \buildrel < \over \sim \;$}
\newcommand{\gtsima}{$\; \buildrel > \over \sim \;$}
\newcommand{\lsim}{\lower.5ex\hbox{\ltsima}}
\newcommand{\gsim}{\lower.5ex\hbox{\gtsima}}

\newcommand{\ahf}{\textsc{AHF}}

\newcommand{\simba}{\textsc{GIZMO-SIMBA}}
\newcommand{\gadgetx}{\textsc{Gadget-X}}
\newcommand{\gadgetmusic}{\textsc{Gadget-MUSIC}}

\newcommand{\Mgas}{M_{\text{gas}}}
\newcommand{\YX}{Y_{\text{X}}}
\newcommand{\YSZ}{Y_{\text{SZ}}}
\newcommand{\Mstar}{M_{\text{star}}}

\newcommand{\AT}{A_{T}}
\newcommand{\BT}{B_{T}}
\newcommand{\ASZ}{A_{SZ}}
\newcommand{\BSZ}{B_{SZ}}
\newcommand{\AX}{A_{X}}
\newcommand{\BX}{B_{X}}

\newcommand{\vpeak}{V_{\text{peak}}}

\newcommand{\thethreehundred}{{\sc The Three Hundred}}
\newcommand{\Tgas}{T_{\text{gas}}}

\usepackage[T1]{fontenc}
\usepackage{ae,aecompl}

\usepackage{newtxtext,newtxmath}

\title[Machine Learning in Galaxy Clusters]{Machine Learning methods to estimate observational properties of galaxy clusters in large volume cosmological N-body simulations}

\author[Daniel de Andres et.al.]{
\parbox{\textwidth}
{Daniel de Andres$^{1,2}$,
Gustavo Yepes$^{1,2}$, 
Federico Sembolini$^{1,3}$,
Gonzalo Martínez-Muñoz$^{4}$, 
Weiguang Cui$^{1,2,5}$,  
Francisco Robledo$^{6,7}$, 
Chia-Hsun Chuang$^{8,9}$,
Elena Rasia$^{10,11}$
}
\\
\\
$^{1}$ Departamento de Física Teórica, M-8, Universidad Autónoma de
Madrid, Cantoblanco 28049, Madrid, Spain\\
$^{2}$ Centro de Investigación Avanzada en Física Fundamental,(CIAFF), Universidad Aut\'{o}noma de Madrid, Cantoblanco, 28049 Madrid, Spain\\
$^{3}$  Equifax Ibérica, Data \& Analytics,  Paseo de la Castellana 259D, Madrid, Spain  \\
$^{4}$  Computer Science Department, Escuela Politécnica Superior,
Universidad Autónoma de Madrid, Cantoblanco, 28049, Spain\\
$^{5}$ Institute for Astronomy, University of Edinburgh, Royal
Observatory, Edinburgh EH9 3HJ, UK \\
$^{6}$ Departamento de Fundamentos del Análisis Económico II, Universidad del País Vasco/Euskal Herriko Unibertsitatea, Barrio Sarriena s/n, \\ 48940 Leioa, Bizkaia, Spain\\
$^{7}$ Laboratoire de Mathématiques et de leurs Applications. Université de Pau et des Pays de l'Adour, Avenue de l'Université, BP 576, 64012 Pau, France\\
$^{8}$ Department of Physics and Astronomy, University of Utah, Salt Lake City, UT 84112, USA \\
$^{9}$ Kavli Institute for Particle Astrophysics and Cosmology, Stanford University, 452 Lomita Mall, Stanford, CA 94305, USA\\
$^{10}$ INAF - Osservatorio Astronomico Trieste, via Tiepolo 11, 34123, Trieste 34123, Italy \\
$^{11}$ Institute of Fundamental Physics of the Universe, via Beirut 2, 34151 Grignano, Trieste, Italy
}
\date{Accepted: receives: in original form: }

\pubyear{2022}

\begin{document}
\label{firstpage}
\pagerange{\pageref{firstpage}--\pageref{lastpage}}

\maketitle

\begin{abstract}

In this paper we study the applicability of a set of supervised machine learning (ML) models  specifically trained to infer observed related properties of the baryonic component (stars and gas) from  a set of features  of dark matter only cluster-size halos. The training set is built  from  \thethreehundred{} project which consists of a series of zoomed  hydrodynamical simulations of cluster-size regions extracted from the 1 Gpc volume {\sc Multidark} dark-matter only simulation  (MDPL2). We use as target variables a set of baryonic properties for the intra cluster gas and stars derived from the hydrodynamical simulations and correlate them with the properties of the dark matter halos from the MDPL2 N-body simulation. The different ML models are trained from this database and subsequently used to  infer the same baryonic properties for the whole range of cluster-size halos identified in the MDPL2. We also test the robustness of the  predictions of the models against mass resolution of the dark matter halos and conclude that their inferred baryonic properties are rather insensitive to their DM properties which are resolved with almost an order of magnitude smaller number of particles. We conclude that the ML models presented in this paper can  be used as an accurate and computationally efficient tool for populating cluster-size halos  with observational related baryonic  properties in large volume N-body simulations making them more valuable  for  comparison with full sky galaxy cluster surveys at different wavelengths. We make the best ML trained model publicly available.

\end{abstract}




\begingroup
\let\clearpage\relax
\endgroup
\section{INTRODUCTION}
\label{sec:intro}

Galaxy  clusters  are  the  largest   gravitationally  bound  objects  of
the  Universe  and  constitute  one  of  the  best  cosmological  probes
to constrain cosmological parameters  of the Universe.
The main component of galaxy clusters is dark matter (DM), which accounts for 85 per cent of the total mass \citep[for a full review see e.g.][]{allen2011cosmological,kravtsov2012formation}. Although the existence of DM is now widely accepted by the scientific community and strongly supported by modern cosmological theories, it has  not been directly detected yet.
To study galaxy clusters, we have therefore to focus on their baryonic component, which represents the remaining 15 per cent of the mass. It  is composed by the hot gas of the Intra Cluster Medium (ICM, around 10-15 per cent of the total cluster mass) and  stars (less than 5 per cent of the mass).\\

Numerical simulations play a fundamental role to study the properties of galaxy clusters. In the simplest scenario, N-body simulations can easily describe the dark-matter component of clusters, which is governed only by gravity; nowadays it is computationally possible to perform very large cosmological simulations, up to a few Gpc$^3$, e.g. {\sc MillenniumXXL} \citep{Langulo2012MillenniumXX}, {\sc MICE} \citep{fosalba2015mice}, {\sc MultiDark} \citep{klypin2016multidark}, {\sc Dark sky} \citep{skillman2014darksky}, {\sc OuterRim} \citep{habib2016hacc}, {\sc FLAGSHIP} \citep{potter2017pkdgrav3}, {\sc Uchuu} \citep{2021MNRAS.506.4210I},  {\sc BACCO} \citep{2021MNRAS.507.5869A} and  {\sc UNIT} project \citep{UNITSIM}, which include thousands of galaxy clusters. Nevertheless, when aiming to describe the baryon component of clusters, due to the complex physics involved in  the  processes  of  cluster  formation,  radiative hydrodynamic  numerical simulations have to be used. These simulations are computationally very expensive so this puts strong limitations to the size of the computational volumes. Examples of state-of-the art of such  simulations are: {\sc Illustris} \citep{vogelsberger2014illustris}, {\sc Eagle} \citep{2015MNRAS.446..521S}, {\sc Horizon-AGN } \citep{2016MNRAS.461.2702C}, {\sc Magneticum} \citep{dolag2016magneticum} or {\sc BAHAMAS} \citep{mccarthy2018bahamas}. Hydrodynamical simulations are essential to calibrate mass proxies and to study the systematics affecting observational measurements. They are also essential to deeply understand the  formation  and  evolution  of  clusters  of  galaxies  and  all  their gas-dynamical effects. For this reason, numerical simulations have been a powerful tool to guide galaxy clusters observations for more than 20 years \citep{evrard199620years,bryan199820years}.

In an ideal scenario one would need to have a large sample of simulated galaxy clusters with enough numerical resolution, both in mass and in the gravity and pressure forces. This high resolution would allow to accurately resolve the internal substructures and to obtain a detailed modelling of the most relevant physical processes. The best way to achieve this would be by simulating large cosmological boxes containing up to tens of thousands of galaxy clusters. 
Unfortunately, due to the large computational effort demanded by these simulations, one needs to find a compromise between their three main components: volume
size, mass resolution and physical processes included. A possible
solution to the computational problems related with scalability of  present-day hydrodynamical codes is to proceed with the so-called `zoom' simulations, such us the MUSIC\footnote{\url{ https://music.ft.uam.es}}  simulation \citep{Sembolini2013}, the {\sc Dianoga } clusters  \citep{planelles2013baryon}, {\sc Rhapsody-G} \citep{wu2015rhapsody}, {\sc MACSIS} \citep{barnes2016macsis},  {\sc Cluster-EAGLE}  \citep{barnes2017eagle}, {\sc hydrangea} \citep{bahe2017hydrangea} clusters and \thethreehundred{} (The300)\footnote{\url{https://the300-project.org} } simulation project \citep{cui2018three}. Zoom simulations are performed mimicking the observations, by creating a catalogue of resimulated galaxy clusters that are extracted from low-resolution N-body simulations. The regions
containing clusters of galaxies are then resimulated at very high
resolution, adding gas physics in the resimulated areas and keeping the rest of the box at low resolution in order to reproduce the same gravitational evolution. 

An alternative approach to  hydrodynamical  simulations to describe the gas and stellar properties of galaxy clusters, is to use Semi-Analytic Models (SAMs), such us {\sc GALACTICUS} \citep{benson2012galacticus}, {\sc SAG} \citep{cora2018SAG}, {\sc SAGE} \citep{croton2016SAGE}  and {\sc GALFORM} \citep{galform}. In this approach, the numerous complex non-linear radiative physical processes associated to the gas-star components  are modelled  using a combination of analytic approximations and empirical calibrations of many free parameters against a set of observational constrains (see e.g. \citet{2006RPPh...69.3101B} for a review). Nevertheless, SAMs are also computationally expensive since most of them  are based on the information provided by merger history of  each individual dark matter halo. A complementary approach is the use of phenomenological models to derive physical properties of the ICM as in \cite{zandanell2018} and \cite{baryonpastingOsato}.
Describing the gas physics  in simulated galaxy clusters requires therefore a big computational effort and impose a compromise between numerical resolution and size of the cosmological volume to simulate.

The main goal of supervised Machine Learning  (ML)  is to generate models that 
  can learn complex relationships between input and output variables from    high-dimensional  data that can later be used to  make predictions on unseen data. 
In this scenario, ML could offer a  powerful alternative to infer some fundamental information on the main  properties (e.g. gas and star masses, gas temperature, etc) of the baryon component of galaxy clusters, without the large computational cost required by hydrodynamical simulations or SAMs. 
Applications of  ML to find a mapping between hydrodynamical and N-body simulations  have  been already presented  in previous works. Firstly, in  \cite{Kamdar2016-zm}, a promising technique to study galaxy formation using numerical simulations and ML was presented; \cite{jo2019machine} estimated galactic baryonic properties mimicking the {\sc{Illustris}}TNG simulation \citep{nelson2019illustris}; \cite{Wadekar_2021neutralhydrogen} generated neutral hydrogen from dark matter; \cite{bernardinibaryonfields} predicted high resolution baryon fields from dark matter simulations; \cite{moews2021hybrid} used hybrid analytic and machine learning model to paint dark matter galactic halos with hydrodynamical properties; \cite{LovellCeagleML} explore the halo-galaxy relationship in the periodic EAGLE simulations, and zoom C-EAGLE simulations of galaxy clusters; and \cite{McGibbonMultiepoch} consider a ML model that is built using the extremely randomised tree (ERT) algorithm and takes subhalo properties over a wide range of redshifts as its input features for galaxy scales.  Recently, The {\sc{CAMELS}} collaboration \citep{CAMELS} has released results from  almost ten thousands simulations (both hydrodynamical and N-body)  with different cosmologies and baryon physical models that are an invaluable tool for training current and future  Artificial Intelligence  algorithms that will be very useful for galaxy formation studies. Unfortunately, given the box sizes, the number of cluster-size objects is poorly represented in these simulations. 

The purpose of this study is to explore the applicability of ML  techniques to generate baryon cluster properties  from DM-only  halo  catalogues  mimicking  the results from \thethreehundred{}  hydrodynamical simulations. More precisely, we use the properties of the cluster-sized halos extracted from parent dark matter only full box simulation MDPL2 as the features  of our dataset. Then we collect several baryon properties of the objects that have been re-simulated with radiative processes and hydrodynamics  as targets (the predicted variables) of the ML models. Our work differs from previous studies in that the baryon properties are extracted from `zoom' MDPL2-based simulations and therefore, we have paired one to  one the objects corresponding to the full N-body only simulations with their hydrodynamical counterparts. As explained below, The300 simulations  corresponds to spherical regions centred on the 324  most massive  clusters found in the MDPL2 box. But there are more  cluster-size halos found within each region with  lower masses. The masses of the  cluster-size  catalogue of hydrodynamical simulated objects we are using ranges from  $\sim 10^{13} \hMsun$ up to $\sim 10^{15}\hMsun$.

The article is structured as follows: In \autoref{sec-2}, we describe how the training dataset is generated using The300 and the MDPL2 simulations. In \autoref{sec-3}, we explain the different ML algorithms used in this work and the training setup. We also study the feature importance and selection of our feature space. In  \autoref{sec-4}, the main results for this work are shown, including an analysis of the performance of the  ML models and their dependence on mass resolution of the simulations. In \autoref{sec-5}, we study the scaling relations extracted from the new ML-generated catalogues and finally in \autoref{sec-6}, we draw our main conclusions and propose possible future studies.

\section{THE TRAINING DATASET}

\label{sec-2}


In order to create the database for training the ML models, we use the MDPL2\footnote{\url{www.cosmosim.org}}  simulation, which has been run using the cosmological parameters measured by the Planck Collaboration \citep{CosmoPlanck}. The MDPL2 simulation consists of a periodic cube volume of comoving length $1 \hGpc$ containing $3840^{3}$ dark-matter particles each with a  mass  $1.5 \times 10^9 \hMsun$.

To build this training dataset,  we need first to identify and extract from  the MDPL2 simulation  the same cluster objects that were used to run zoomed The300 hydrodynamical simulations. We then select the main properties of the dark matter clusters  and associate them with the    baryonic properties extracted from their The300  hydrodynamical counterparts.

\subsection{MDPL2: Dark Matter input variables}

 In order to identify the dark matter halos and measure their internal properties in the MDPL2 N-body simulation we have used the {\sc Rockstar} halo finder \citep{Rockstar}, complemented  with additional information based  on the halo  mass accretion history from the {\sc Consistent Halo Merger Trees}  analysis \citep{2013ApJ...763...18B}. We  have extracted a total of 26 relevant physical {\sc Rockstar + Consistent Trees} variables\footnote{More information regarding the selection of { \sc Rockstar} variables can be found in Appendix \ref{appendixA}} (masses at different radii, velocities, symmetry factors, properties related with mass accretion history, etc) to create our dark matter catalogue.  In addition, we have also considered the scale factor $a(z)$ of clusters as an input variable. Furthermore, we have introduced a cut-off in halo mass such that $\log (M /(\hMsun ) )
 \geq 13.5$ and redshift $\leq 1.03$.

In \autoref{fig:correlation}, we show the Spearman correlation matrix of the 26 {\sc Rockstar} variables and the scale factor $a(z)$. These variables are ordered using a hierarchical clustering algorithm based on Ward’s linkage on a condensed distance matrix. We used the Python  implementation of this algorithm from  {\sc SciPy} \citep{2020SciPy-NMeth}. We can easily identify 5 groups in the correlation matrix. The first group  (variables 0 to 12) corresponds to masses and velocities at
different radii. In a second group,  different ellipticity shape  factors (from 13 to 16) are included. Variables from 17 to 21 corresponds to the scale radius, the ratio between the kinetic and potential energy and the offsets between density peak  and centre-of-mass, which are directly related to the dynamical state of the cluster halos. The next group of variables (22 and 23) correspond to the dimensionless spin parameters  of the cluster. Finally, variables from 24 to 26 represent the scale factor (redshift)  and the time evolution of mass accretion. As can be seen in the figure, feature variables inside the same block are strongly correlated among them and they are weakly, or not correlated to variables inside other blocks. This might imply that selecting more than one feature belonging to the same block could not add any new predictive information. This is studied in detail in section \autoref{sec-3}. A more detailed description of the selected feature variables can be found in the Appendix \ref{appendixA}.

\subsection{The300: baryonic output variables}
 Subsequently, for a subset of the MDPL2  cluster halos,  we need to have their baryonic properties. For this purpose, we have used the results of The300 project, which has  re-simulated spherical regions of radius $15 \hMpc$ centred around the 324 most massive clusters  found in the  MDPL2 simulation at $z=0$.  These regions were then mapped  back to the initial conditions and their  particles  were  split into gas and dark-matter, while the rest of the particles in the remaining box were re-sampled into different levels of lower resolution and  larger masses. With this zoom-in  technique, we ensure that the subsequent gravitational evolution  will reproduce the same objects in the high resolution area while we minimise the effects of contamination of low resolution particles from external regions due to  mass segregation.  In any case, we checked that all the clusters used in this work  are free from contamination of low mass resolution particles at least  within their virial radii.
 
 The300 project has produced different versions of hydrodynamical simulations from these initial zoomed conditions which include different baryonic physics modules:  radiative cooling, star formation  and Supernovae Feedback using the    {\gadgetmusic} SPH+TreePM code  \citep{Sembolini2013} and   newer versions that include  feedbacks from Super Massive Black Holes: {\gadgetx} \citep{Murante2010, Rasia2015}, {\simba}  \citep{Dave2019,Cui2022}.

 However, in this work, we  only  make use of  the \gadgetx\ runs.  The halos  in these simulations are identified and analysed  with the {\sc Amiga Halo Finder} (AHF) \citep{AHF}, which is more suitable than {\sc Rockstar}  for simulations with multiple particles species (i.e. dark matter particles, gas, stellar particles and Black Holes).  From the information contained in  the AHF catalogues, we have collected the following baryon properties:

\begin{itemize}
    \item 
The total  \textit{gas mass } $\Mgas$ inside a spherical volume with a mass  density that is   500 times larger  than the critical density of the Universe at each redshift. The radius of this sphere is denoted as $R_{500}$. 
\item
The \textit{Stellar mass } $\Mstar$ inside $R_{500}$. 
\item
The \textit{gas temperature} $\Tgas$ computed as the mass weighted temperature,  inside $R_{500}$
\begin{equation}
    T = \frac{\sum_{i\in R_{500}} T_{i}m_{i}}{\sum_{i\in R_{500}}m_{i}},
\end{equation}
where $T_{i}$ and $m_{i}$ are respectively the temperature and mass of the gas particle. We additionally made a cut of $T>0.3 \text{keV}$ to exclude low-temperature gas particles. 
\item
The \textit{X-ray Y-parameter} $\YX$ defined  as $\Tgas \times \Mgas$, which related with the total thermal energy of the gas and it has been shown  that it is a good proxy of the total cluster mass \citep{2006ApJ...650..128K}.   Note that this quantity can be derived from others. However, we prefer to treat it as an independent target, i.e the ML models are  also trained to predict $\YX$ as one of the target variables.
\item
The \textit{integrated Compton-y parameter} $\YSZ$ over $R_{500}$ given by the Sunyaev-Zel’dovich (SZ) effect \citep{SZeffect}. Particularly, the integrated value $\YSZ$ is computed  from Compton-y parameter maps estimated as in the following:
\begin{equation}\label{eq:defy}
    y= \frac{\sigma_{\text{T}}k_{\text{B}}}{m_{\text{e}}c^{2}}\int n_{\text{e}}T_{\text{e}}dl \text{ ,}
\end{equation}
where $\sigma_{\text{T}}$ is the Thomson cross section, $k_{\text{B}}$ is the Boltzmann constant, $c$ the speed of light, $m_{\text{e}}$ the electron rest-mass, $n_{\text{e}}$ the electron number density, $T_{\text{e}}$ is the electron temperature and the integration is done along the observer's line of sight. Assuming  $dV=dAdl$, Eq.(\ref{eq:defy}) is computed in our simulated data as in \cite{Sembolini2013} and \cite{le2015testing}:
\begin{equation}
    y = \frac{\sigma_{\text{T}}k_{\text{B}}}{m_{\text{e}}c^{2}dA}\sum_{\text{i}}T_{\text{i}}N_{\text{e,i}}W(r,h_{i})\text{ .}
\end{equation}
\end{itemize}
Note that here we have used the number of electrons in the gas particles $N_{\text{e}}$ given that $n_{\text{e}}=N_{\text{e}}/dV$. Moreover, $W(r,h_{\text{i}})$ is the same SPH smoothing kernel as in the hydrodynamical simulation  with smoothing length $h_{\text{i}}$. The $y$-maps are generated with the centre on the projected maximum density peak position of the halo. Each image has a fixed angular resolution of $5^{\prime\prime}$ that is extended to at least $R_{200}$ in all the clusters. The clusters at $z=0$ are placed at $z=0.05$ to generate the mock images while the clusters at higher redshifts simply use its original  value from the simulations. We then integrate the Compton-y map up to $R_{500}$ using only the z-plane projection. Since the dataset is large, the effect of projections is negligible. Note that this approach of estimating $\YSZ$ gives us the cylindrical Compton-y parameter $Y_{500,\text{cyl}}$. In practice, $Y_{500,\text{cyl}}$ is related to the corresponding spherically integrated value $Y_{500,\text{sph}}$ such that $Y_{500,\text{cyl}}/Y_{500,\text{sph}}\simeq 1.2$ as given by the \cite{Arnaud2010} pressure profile and also  compatible with results from numerical simulations \citep{Sembolini2013}.

Note that star forming gas particles with $SFR > 0.1 M_\odot\mbox{yr}^{-1}$ are also excluded in the calculation of $\Tgas$ $\YX$ and $\YSZ$. This is commonly adopted in simulations with a multiphase subgrid  physics model in which gas particles are composed of hot and cold gas components (see e.g. \citet{Borgani2004}). These star forming gas particles are poorly modelled as  they are mostly made by the  neutral cold  gas component which should not emit any X-rays or contribute to the total electron thermal pressure.\footnote{We have verified that changing to $SFR > 0$ makes no difference in the results due to a very small number of star forming gas particles in galaxy clusters\citep[see][]{Li2020}.}

\subsection{The Final Training Dataset}

 After defining our input and output variables, we finally match one-by-one the clusters between the two simulations  that fulfil these two conditions for   the  relative shifts between the cluster centres and the halos mass differences:
 
 \begin{equation}
     \text{distance}(\text{C}_{\text{MDPL2}},\text{C}_{\text{The300}})<0.4\times R_{200,\text{The300}}\text{ ,} 
 \end{equation}
\begin{equation}
     \left \lvert\frac{M_{\text{MDPL2,200}}}{M_{\text{The300},200}}-1\right \rvert<0.1 \text{ .}
\end{equation}

 Here, $C_{\text{MDPL2}}$ and $C_{\text{The300}}$ stand for the centre of mass of the clusters while $M_{\text{MDPL2,200}}$ and $M_{\text{The300},200}$ stand for the mass inside a sphere of radius $R=R_{200}$ for each simulations (between DM only {\sc Rockstar}  catalogue and the {\ahf} catalogue respectively). Due to both the baryon effect \citep[see][for example]{Cui2012,Cui2014} and to different algorithms used by the halo finders, it is not possible to determine with all certainty that all the halos are exactly matched. Notice that the centre difference can be as high as $0.4 R_{200}$. However, with this restrictive selection criteria, only the true/very close counterparts are selected. In this way, we finally provide the baryon properties for the matched MDPL2 clusters using the corresponding The300 objects.
 
\begin{figure}
\includegraphics[width=0.46\textwidth]{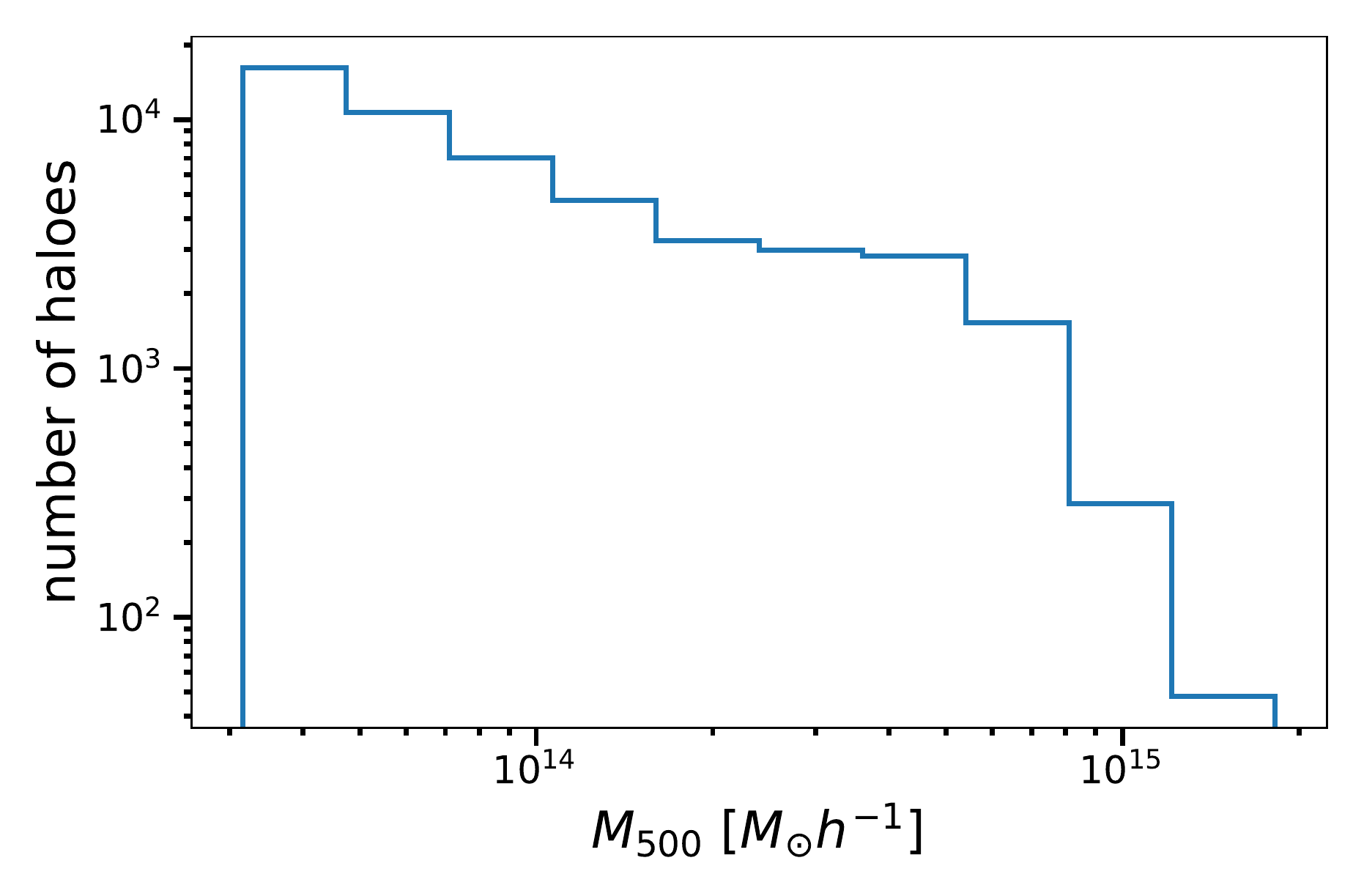}
\caption{ Mass distribution of the 
The300
Galaxy clusters analysed in this work}
\label{fig:mhist}
\end{figure}

\begin{figure}
\includegraphics[width=0.46\textwidth]{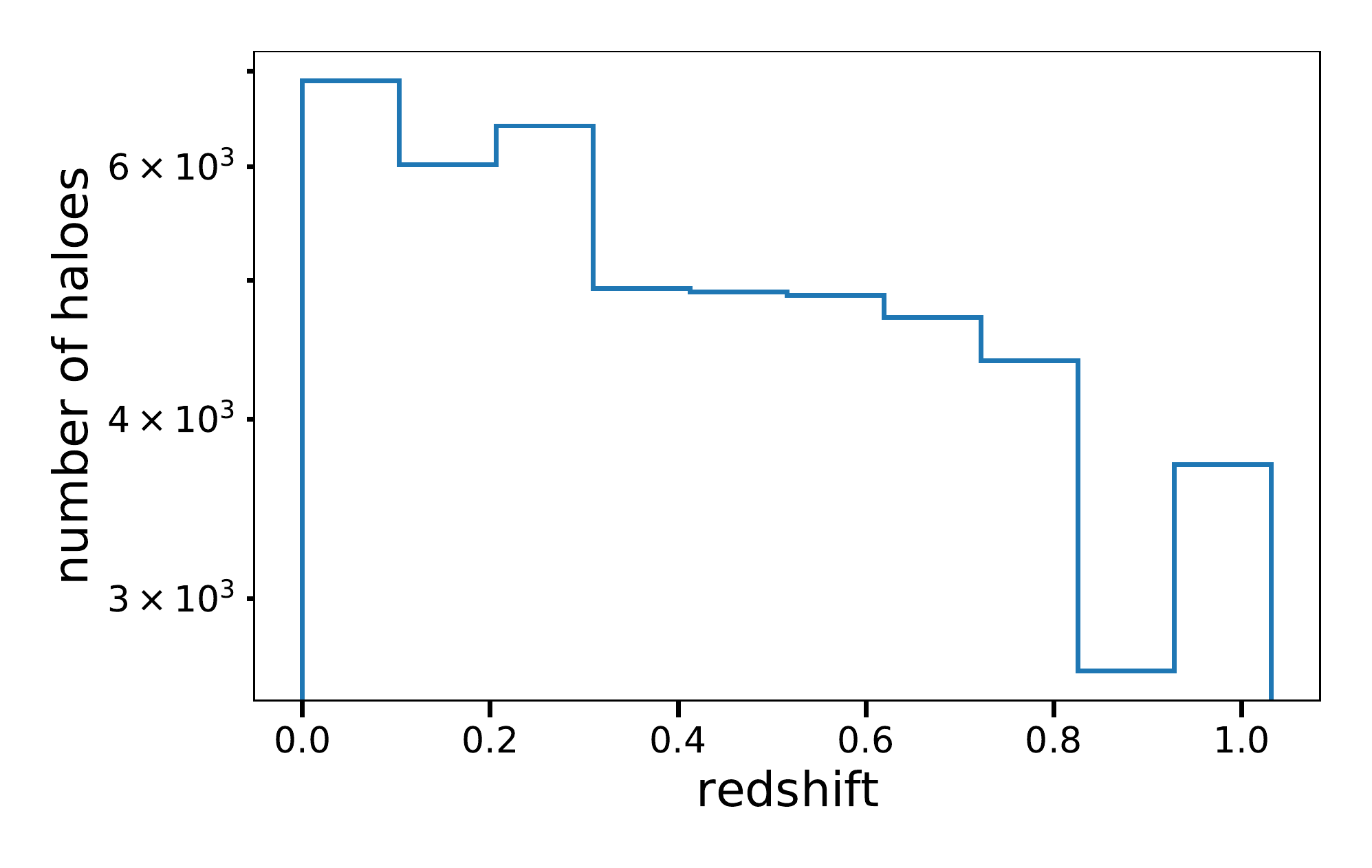}

\caption{Redshift distribution of  The300 Galaxy clusters analysed in this work}
\label{fig:zhist}
\end{figure}

 After this procedure, our dataset is finally composed of 49540 different objects.  Note that all the 33 halo catalogues available  from  $z=0$ to $z=1.03$ in the two simulations have been considered. Only 1264 objects correspond to clusters at $z=0$, the rest of them are the progenitors of the same objects at different redshifts.  The number of objects as a function of their mass and redshift can respectively be found in \autoref{fig:mhist} and \autoref{fig:zhist}. Our final dataset is composed of 27 DM input variables and 5 baryon output variables. These are the features and targets which are used for training and testing the ML algorithms described in the next section.

\begin{figure*}
\includegraphics[width=0.9\textwidth]{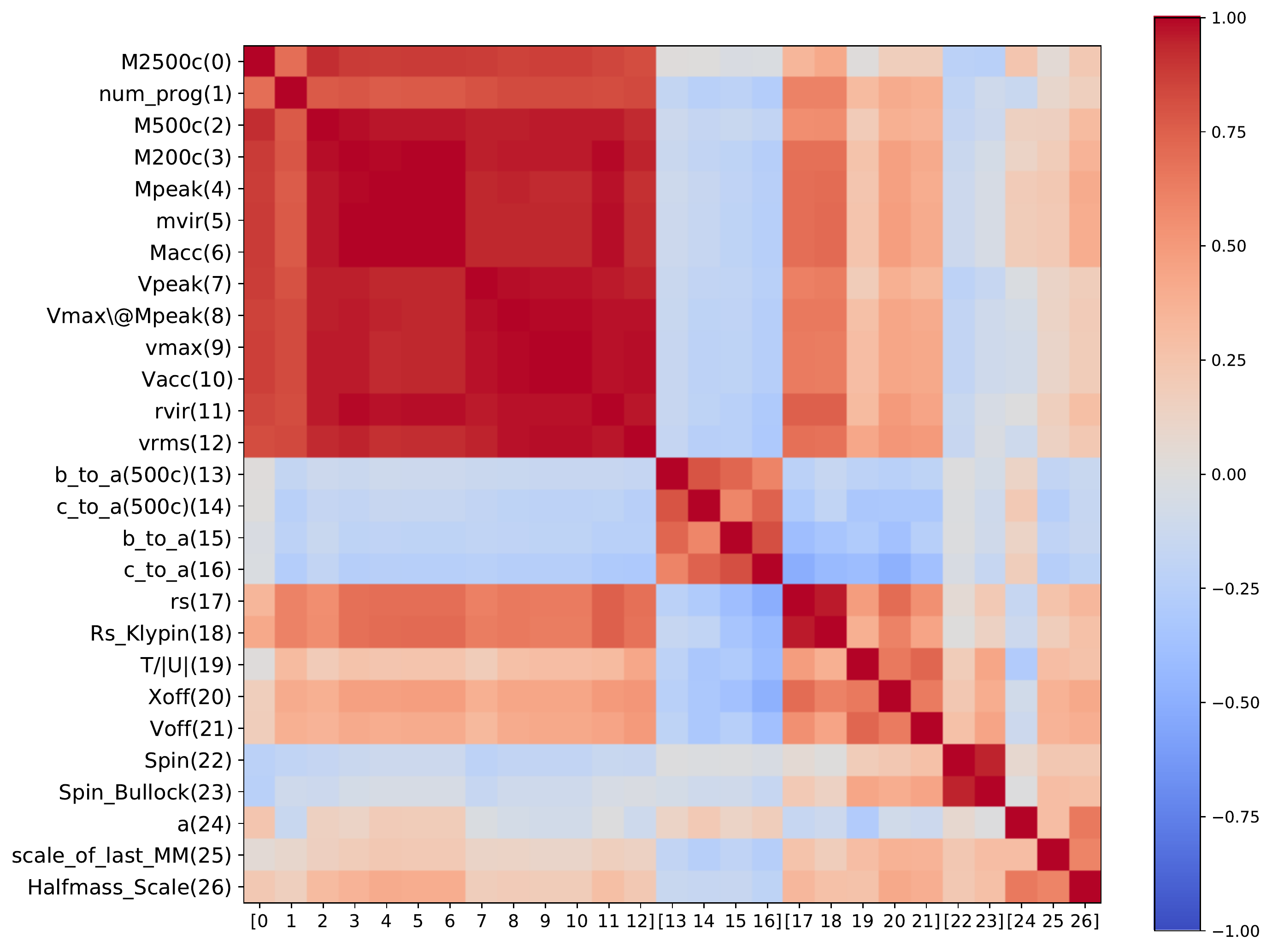}
\caption{Spearman correlation coefficient matrix for the  (feature) variables of the {\sc Rockstar} identified clusters. The variables are organised in different blocks according to their correlation values. Variables for each block are denoted in the x-axis in brackets: [1,...,12] are mass and velocity variables, [13,...,16] correspond to ellipticity, [17,...,21] are related to the dynamical state of the cluster, [22,23] represent dimensionless spin  parameters and [24,25,26] are related to the scale factor and time evolution of mass accretion. Note that this matrix is symmetric with respect to the diagonal. Each variable description can be found in Appendix \ref{appendixA}.}

\label{fig:correlation}
\end{figure*}

\section{MACHINE LEARNING ALGORITHMS: DESCRIPTION AND TRAINING }
\label{sec-3}

In this section, we first describe the machine learning algorithms used in this work and the  training setup. Then, we study the importance of our feature variables in order to reduce the dimensionality of our dataset.

\subsection{Machine Learning Algorithms and Training Setup}
In order to estimate the baryon properties of the dark matter only clusters, several effective supervised machine 
learning methods have been employed. We particularly focus on  four methods: random 
forest \citep[RF;][]{breiman01rf}, extreme gradient boosting \citep[XGBoost; ][]{chen16XGBoost}, dense Neural Networks or Multilayer Perceptron
\citep[MLP; ][]{schmidhuber2015deep} and Natural Gradient Boosting for Probabilistic Prediction \citep[NGBoost;][]{NGBoost2019}. RF and XGBoost have shown to be among the best 
machine learning methods for tabular data (i.e. without a known grid-like topology, such as images) 
\citep{fernandez14comp,bentejac21comp,comparison2017zhang}. Convolutional deep neural 
network models have shown spectacular performance for image-based and structured data 
in general \citep{schmidhuber2015deep}. However, for tabular data, as is the case of
this study, their performance is poor \citep{comparison2017zhang}. Notwithstanding,
deep dense networks can perform well in these scenarios, so we will also consider
these models.

Random Forest and XGBoost are ensemble models  composed of decision trees. During training, these algorithms build hundreds of decision trees from a single training dataset. The process for building these trees in random forest and XGBoost, is based on quite different ideas. Although, the objective in both cases is to build decision tree models that complement each other in order to obtain a classification/regression model better than any of its parts \citep{dietterich98machinelearning}. 

Random forest rely on stochastic techniques to generate many random solutions to the problem at hand. In order to generate each single tree, the random forest algorithm first generates a new dataset by extracting at random $N$ instances of the training data of size $N$ with replacement (i.e. bootstrap sample). This bootstrap sample is used to train a decision tree in which the best split at each node of the tree is selected from a random subsample of features of the data. Generally, the size of the random subset of features is of the order of $\sqrt{D}$ or $\log_2(D)$, with $D$ the number of features of the problem. The final output of the random forest for a given instance is obtained as the mode or mean of all trees for classification and regression respectively. In addition, since the randomisation process to build the trees is independent, the process of building a random forest can be easily parallelised.

On the other hand, XGBoost relies mainly in a gradient descend approach although it  also incorporates stochastic techniques to further increase 
its performance. XGBoost is an additive model based on Gradient Boosting. 
The output of an additive model is the sum of the outputs of its components.
In order, to create this ensemble, regression trees 
are trained sequentially to approximate the gradient of the loss function of the data in the previous iterations. Hence, each new tree learns the {\it remainder} of the concept not learned in previous steps. XGBoost also includes a penalisation term in the number of leaves of the trees to avoid over-fitting. In addition, XGBoost incorporates random feature selection, bootstrap sample and several other randomisation features.

In order to perform a fair comparison among algorithms and also to obtain good estimations of the performance of the different algorithms, we carried out the following experimental procedure based on K-fold cross-validation and grid-search. K-fold cross-validation consist in splitting the data into K disjoint sets of approximately equal size and then to use iteratively $K-1$ sets for training the model and the remaining set for validation.
The main experiment is performed using the same 10-fold cross-validation for the prediction of the five baryonic properties analysed in this study using the {\sc Rockstar} halo catalogue from The300 hydro clusters. The steps for each of the 10 partitions of the cross-validation are:
\begin{itemize}

\item[(i)] Find the best hyper-parameters of each of the tested algorithms: RF, XGBoost and MLP. For that, a grid-search with 5-fold cross-validation within the train dataset only was performed. The values for the grid of hyper-parameters are shown below; 
\item[(ii)] The best set of hyper-parameters for each method were used to train a single model using the whole training set;
\item[(iii)] The models were validated using the test set; 
\end{itemize}
In order to generate dark matter only     halo catalogues with hydrodynamic properties, the 10 trained  models from each of the 10-folds of the cross-validation were used. The hydrodynamic features of each halo are then computed as the average of the inferred values from  these 10  models.

For the grid search the set of values of the tested hyper-parameters for each of the analysed methods are:

\begin{itemize}

\item \textit{Random Forest}:

\begin{itemize}
    \item The number of trees in the forest: `n\_estimators'=[100,500]
    \item the number of features to consider when looking for the best split `max\_features' : [ round($D^{1/2}$), round(`$\log_{2} (D)$')] 
\end{itemize}

\item \textit{XGBoost:} 

\begin{itemize}
    \item `n\_estimators'= [100,500]
    \item Maximum depth of a tree:
    
    `max\_depth'= [5,6,10,14,15,16,20]
    \item Minimum loss reduction required to make a further partition on a leaf node of the tree: 
    
    `gamma' = [0,0.1,0.2,0.3,0.4,0.5,0.6,0.7,0.8,0.9,1]
    \item Step size shrinkage used in update to prevent overfitting: 
    
    `eta' = [0,0.1,0.2,0.3,0.4,0.5,0.6,0.7,0.8,0.9,1]
\end{itemize}

\item \textit{MLP}: 

\begin{itemize}
    \item `hidden\_layer\_sizes' = [(8,),(20,),(100,),(8,8),(8,20,8),\\ (20,20,20),(100,100,100),
 (20,20,20,20),(100,100,100,100)] 
 \item  `activation'=`relu',
 \item `solver'=`adam',
 \item `learning\_rate'=$10^{-4}$
\end{itemize}
\end{itemize}

Furthermore, MLP has been trained for 500 epochs or until the training loss is constant during 20 epochs. For more information of these hyper-parameters, we refer the reader to the Python libraries used throughout this work: for RF and MLP we have used  scikit-learn\footnote{\url{https://scikit-learn.org}} \citep{scikit-learn} and for XGBoost its own library\footnote{\url{ https://github.com/dmlc/xgboost}}. For the hyper-parameters not considered in the search grids, their defaults values were used.

In order to train these models the mean squared error of the logarithmic values of the targets (logarithmic MSE) was used as the loss function:
\begin{equation}\label{eq:lossf}
\text{MSE}=\frac{1}{n}\sum_{i=1}^{n}(\log y_{\text{true},i}-{<\log y}_{\text{pred},i}|x_{i}> )^{2}\hspace{1mm},
\end{equation}
where $y_{\text{true},i}$ is the true value of the target extracted from the The300 simulation and <${\log y}_{\text{pred},i}|x$> is the predicted target's value by our model.  Note that, since the model is trained with $\log y$ as targets, then the prediction of the model is directly the logarithm of the given target. In addition, $n$ corresponds to the number of objects in the dataset (e.g. train, validation and test set) where the MSE is computed.
For numerical reasons, we have also used the logarithmic value of the features during the training process. 

Also note that these models infer only one prediction for every input value $x_{i}$. These predictions are considered as the mean predictions without their statistical uncertainties. Moreover, \cite{StiskalekScatter2022} showed that by accounting for a proper modelling of uncertainties ML models can successfully mimic the statistics of the data, i.e. not only the mean but also the  scatter. Moreover, modelling uncertainties using ML algorithms is a topic of recent studies  \cite[e.g][]{Uncertainties1,Uncertainties2,Uncertainties3,Uncertainties4,deAndres2022Planck}

In order to address this issue we apply a generalisation of gradient boosting for probabilistic modelling: NGBoost. NGBoost is a gradient boosting algorithm as XGBoost, but that is based on assuming a particular parametric ($\theta$) probability distribution $P_{\theta}(y|x)$. Thus, the loss function can be written in terms of these parameters as the negative log-likelihood:
\begin{equation}
\mathcal{L}(\theta,y)=-\log P_{\theta}(y).    
\end{equation}
In our work we assume a Gaussian distribution with parameters mean $\mu_{i}=<{\log y}_{\text{pred},i}|x_{i}>$ and standard deviation $\sigma_{i}$. Therefore, the last equation can be written as
\begin{equation}
\mathcal{L}_{\text{NGBoost}}(\theta,y)= \frac{1}{n}\sum_{i=1}^{n}\left( \ln \sigma^{2}_{i} + \frac{(\log y_{\text{true,i}}-<{\log y}_{\text{pred},i}|x_{i}>)^{2}}{2\sigma^{2}_{i}}\right).
\end{equation}
Note that for every input example NGBoost predicts a mean value $\mu$ and its scatter which is given by the width of the Gaussian $\sigma$. NGBoost also implements the generalised natural gradient when minimising the loss function.

In order to tune NGBoost, we follow a procedure similar to the one proposed in the original paper. For every K-fold, we find the best value of the number of estimators by cross-validation. Moreover, for all the experiments the base learners are decision trees with a maximum depth of three levels and the learning rate is set to 0.01. In summary, the hyper-parameters for NGBoost are:
\begin{itemize}
    \item \textit{NGBoost}
    \begin{itemize}
    \item `n\_estimators'= [best out of a maximum of 3000]
    \item `max\_depth'= 3
    \item `learning\_rate'= 0.01
\end{itemize}
\end{itemize}

All other hyper-parameters were set to their default values. Note that NGBoost is the only model that can generalise the intrinsic scatter:
\begin{equation}
    (\log y|x) = <(\log y|x)> + \epsilon, 
\end{equation}
where $\epsilon$ is the noise due to the scatter.
Previous models (RF,XGBoost, MLP) are only designed to infer the mean value of $<\log y|x>$. Moreover, NGBoost predictions on unseen data are computed by averaging over the 10 models' distributions corresponding to our 10 different K-folds:
\begin{equation}
    \mu =  \frac{1}{10}\sum_{k=1}^{10}\mu_{k},
\end{equation}
\begin{equation}
    \sigma^{2} =  \frac{1}{10}\sum_{k=1}^{10}\sigma_{k}^{2}+(\mu_{k}-\mu)^{2}.
\end{equation}

\subsection{Feature importance and selection}
Although machine learning models can generalise complex functions, 
 generally, it is not trivial to interpret their decisions. In fact, they are often referred as black box estimators \citep[e.g.][]{BARREDOARRIETA202082}.
Therefore, it is of great value to be able to inspect what is the learnt relation between features and targets given a particular model. One such inspection technique is feature importance. Particularly, feature importance is a family of techniques that assigns a score $F_s (x,y)$ to each input features $x$ depending on how useful they are when it comes to predicting a particular target $y$.
Furthermore, feature selection is a family of techniques  
that aim at getting  rid of non-informative variables from a model \citep[e.g.][]{kuhn2013applied}. In this section, we use a feature importance algorithm to determine what features are more relevant and therefore, reduce the dimensionality of our 27-dimensional input space. 

 One commonly used algorithm to estimate feature importance for ensembles of decision trees (such as RF and XGBoost) is {\it Permutation Importance} \citep{breiman01rf}. In this algorithm, the importance of each feature is estimated
as the decrease of the model score when the values of a feature are randomly shuffled. This technique, however, fails when correlated features are present in the dataset \citep{Altmann20101340}. A second shortcoming of this algorithm is that it only considers the importance of individual features.

Other technique is the use of forest of trees to evaluate the importance of features computed as the mean and standard deviation of accumulation of impurity decrease within each tree \citep{breiman01rf}, which for regression is the variance reduction. In random forest, internal node features are selected with some criterion, or loss function. We can then measure how on average each feature decreases the criterion in the splits of the forests. Nevertheless, this technique also fails due to the fact that our features are highly correlated, and it is also known to be biased in favour of variables with many possible split points   \citep[e.g][]{nembrini2018revival}.

Instead, we use the \textit{Greedy Search Feature Importance Algorithm} (GSFIA,  see for example \citet{ferri1994comparative}. This technique  considers the importance of the combination of features and not only the individual feature importance. It works iteratively by selecting  and evaluating one variable at a time until all features are ordered from the most to the least relevant. The algorithm works with a list of selected variables, $L$, initially empty, $L=[]$, and a pool of possible variables to be selected, $P$, initially containing all $D$ variables of the problem, $P=set(x_1, \dots , x_D)$. Then, a procedure is repeated $D$ times in which, at each step, one variable from the pool $P$ is selected and moved to the list $L$. In the $k-\text{th}$ step of the loop the procedure creates $|P|$ models trained on  all of the features in $L$ plus one feature from $P$. The model that minimises the MSE identifies the most important variable from $P$ in combination with the variables in $L$. This variable is then removed from $P$ and appended to the list $L$. At the end of the algorithm, all variables of the problem are sorted by importance in list $L$ together with the loss function associated with them. GSFIA it is depicted using pseudocode in \autoref{alg:gsfia}.

\begin{figure}
\includegraphics[width=0.46\textwidth]{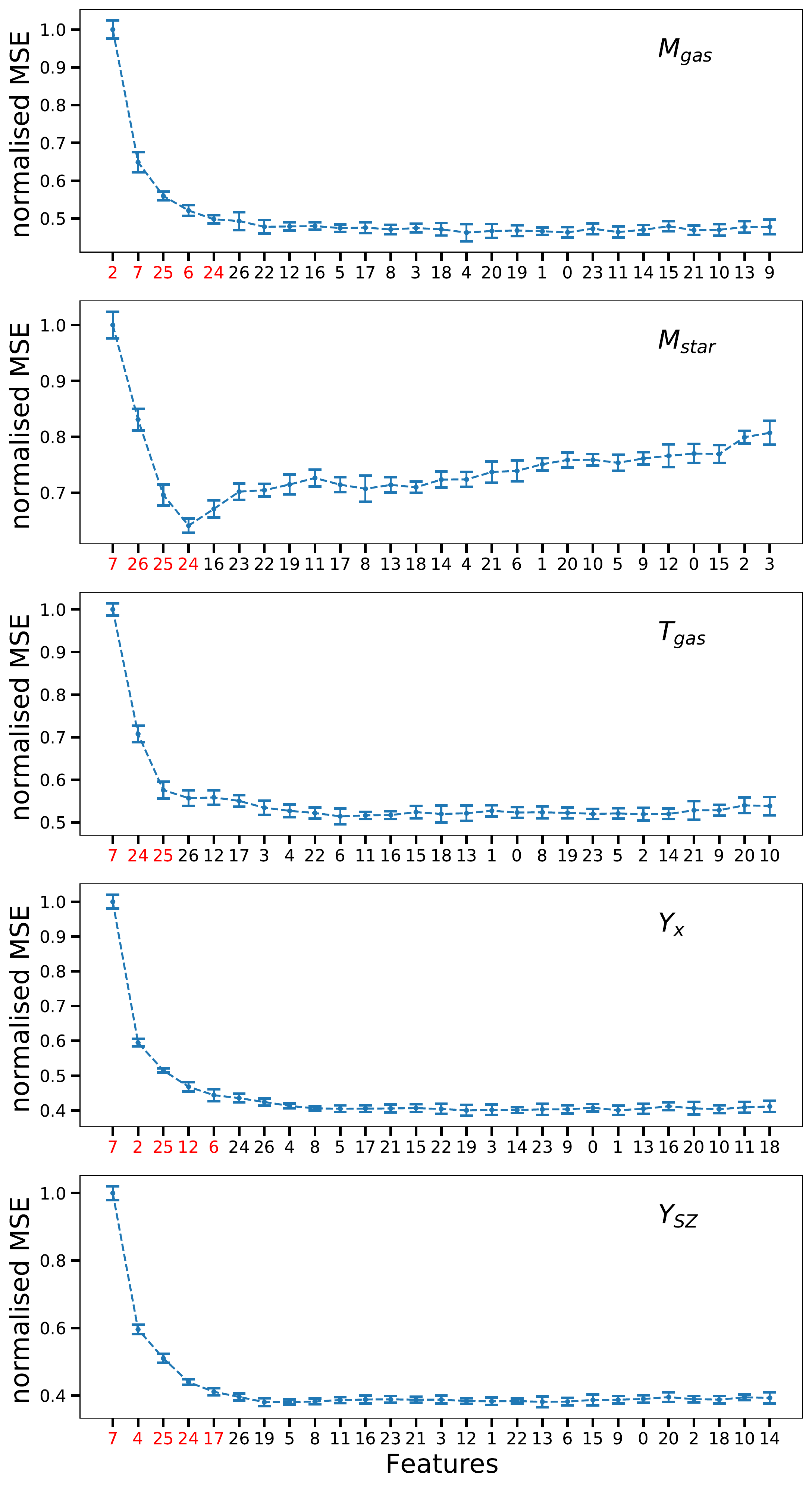} 
\caption{Normalised MSE (y-axis) given by GSFIA (\autoref{alg:gsfia}) as a function of  DM-only variables described in Appendix \ref{appendixA} ranked by  feature importance in descending order. From top to bottom we show our results for different targets: $\Mgas$, $\Mstar$, $\Tgas$, $\YX$ and $\YSZ$. Blue dashed lines represent the average value of the normalised MSE for 10 different k-folds and error bars correspond to the standard deviation. The selected features for each target are highlighted in red and shown in \autoref{table:important_features}.}
\label{fig:importance}
\end{figure}

With this algorithm, we can  
define the feature importance score $F(x,y)$ as follows: 
\begin{itemize}
    \item Run GSFIA to rank  all features from the most to the least important variables and save the corresponding value of  MSE.
    \item The score is then defined as the MSE of every iteration normalised to the corresponding value of the first iteration.
\end{itemize}
Note that the normalised MSE will be 1 for the first feature, and will decrease progressively as we consider more features until it converges to a minimum value. It could happen that after including several features, the normalised MSE increases as more features are included (see \autoref{fig:importance} for the  case of $M_{\text{star}}$). This indicates that the last features included do not improve or even degrade the performance of the model.

\begin{algorithm}[]
\SetKwInOut{Input}{inputs}
\SetKwInOut{Output}{outputs}

\Input{x = features dataset; y = target dataset}
\Output{L = organised list of features according to their importance degree ; score = normalised MSE of every element in L}
 L = [empty list]\;
 score = [empty list]\;
 P = [$x_{1}$, ... ,$x_{D}$]\;
 \For{i in 1...D}{
 loss = zeros(length(P))\;
 j = 1\;
 \While{j<=length(P)}{
    dataset = L+P[j] \# sum of lists \;
    model.train(dataset,y)\;
    loss[j]=model.MSE\;
    j=j+1\;
    }
 indx = argmin(loss)\;
 L.append(P[indx])\;
 P.drop(indx)\;
 score.append(loss[indx])\;
 }
 score = score / score[0]\;
return L, score
 
\caption{Pseudo code of Greedy Search Feature Importance Algorithm}
\label{alg:gsfia}
\end{algorithm}
 
The algorithm was run using random forest as model (line 9 of \autoref{alg:gsfia}). In addition, due to the randomness of the ML model, the inner loop of the algorithm was repeated 10 times in order to reduce the variability of the results.
In Figure~\ref{fig:importance}, the average of the normalised
 MSE and its standard deviation are shown for the different targets considered. In the horizontal axis, the final order for the feature variables is shown. Variables in red colour are the reduced set of features that will be considered for further analysis. These features are summarised in \autoref{table:important_features}.

 As shown in \autoref{table:important_features}, we expect that the selected variables generally come from different correlation blocks, as shown in \autoref{fig:correlation}.  This is so, since variables from the same block are correlated and once the algorithm chooses one feature, it skips using variables with the same information.  However, this is not always the case (e.g. variables 2, 6 and 7 are selected for a couple of targets). This can be explained since the correlation between those variables is high, but it is not 1. Hence, for our case the marginal information that a second variable inside a correlated block gives, is higher than that given by other variables. As far as the meaning of selected variables is concerned, we can distinguish two different important blocks in the correlation matrix: The mass and velocity block (the first block from 0 to 12), and the time evolution block (from 24 to 26). The conclusion of this analysis is that the rest of the blocks are redundant or contribute little to the estimating baryon properties, i.e. the ellipticity block (from 13 to 16), the dynamical state block (from 17 to 21) and the spin block. Moreover, masses and velocities are the most important features for estimating baryon properties while the variables associated with the time evolution of the mass accretion into halos  play a secondary role in the regression algorithms. The redundant role of the ellipticity variables can be explained by taking into consideration that we are estimating integrated quantities from the particles  within  spheres of radius $R=R_{500}$, regardless of the  shape of their 3D distributions.
 
 Note that, we combine data from different redshifts as our training and test samples. We do not think that the evolution of these baryon properties will affect our results because (1) as shown in \cite{Cui2022}, these quantities in \gadgetx\ simulations  hardly   depend on redshift, especially at $z\lesssim1$ \citep[see also][for example]{Truong2018}; (2) we also include the scale factor as a feature variable in the training. If there were a clear redshift dependence on any target variable, the scale factor feature  would show a higher  contribution. However, as shown in \autoref{fig:importance}, the scale factor  contributes only weakly to the normalised MSE.

Furthermore, we have to highlight that although we have used 
Random Forest for the GSFIA, other Machine Learning algorithms might also  be used. However, GSFIA is computationally expensive given the fact that its computing time increases with the number of features $D$ as $O(D^{2})$.  Therefore, we prefer to use RF because it is computationally more efficient and it does not have as many hyper-parameters to tune in. Consequently, this choice might introduce a bias given the fact that a particular model is being used for the selection of the important variables. However, in the next section we will show that this particular selection of variables yields  similar performance for the different ML algorithms considered throughout this work.

\begin{table}
\centering
\caption{Lists of selected DM-only features for the different targets after applying GSFIA (\autoref{alg:gsfia}).}
\label{table:important_features}
\begin{tabular}{cc} 
\toprule
\textbf{target } & \textbf{Important features }                                   \\ 
\hline
$\Mgas$          & M500c(2), Vpeak(7), scale\_of\_last\_MM(25), Macc(6), a(24)    \\
$\Mstar$         & Vpeak(7), Halfmass\_Scale(26), scale\_of\_last\_MM(25), a(24)  \\
$\Tgas$          & Vpeak(7), a(24), scale\_of\_last\_MM(25)                       \\
$\YX$            & Vpeak(7), M500c(2), scale\_of\_last\_MM(25), vrms(12),Macc(6)  \\
$\YSZ$           & Vpeak(7), Mpeak(4), scale\_of\_last\_MM(25), a(24), rs(17)     \\
\bottomrule
\end{tabular}
\end{table}

\section{RESULTS}\label{sec-4}

\begin{figure}
\includegraphics[width=0.48\textwidth]{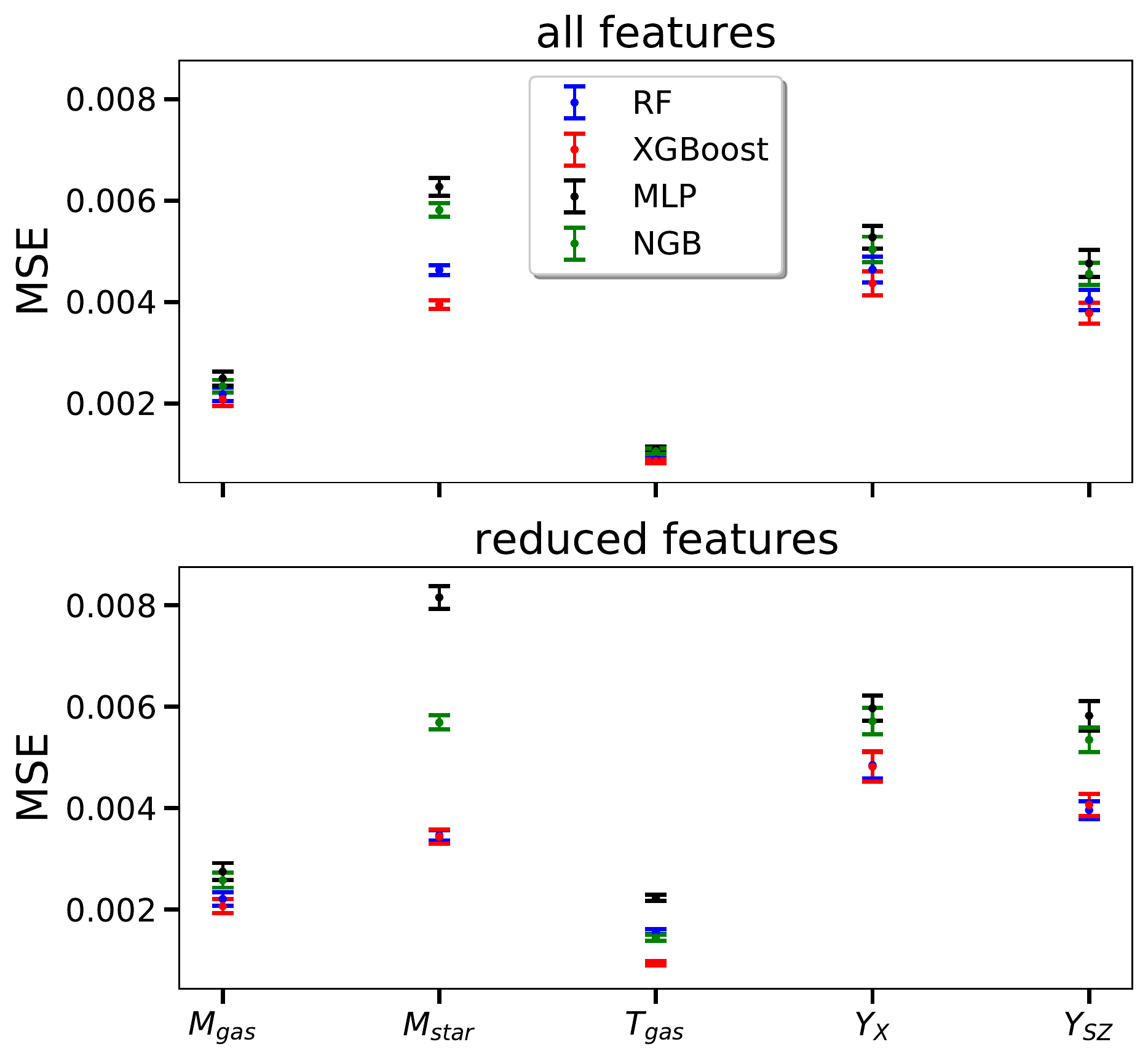}
\caption{The logarithmic MSE defined in Eq.(\ref{eq:lossf}) for the four ML models considered: RF in blue, XGBoost in red, MLP in black and NGB in green colour. The x-axis indicates the different baryon targets. The points with error bars represent the mean and the standard deviation of the logarithmic MSE for the test set using 10 different k-folds. The top panel corresponds to training the ML models using all the DM-only variables shown in \autoref{fig:correlation} and in the bottom panel the algorithms are trained with the reduced set of features listed in  \autoref{table:important_features}.}
\label{fig:rMSE}
\end{figure}
In this section, we first study what machine learning algorithm is of higher quality for our particular dataset and study the accuracy of our model predictions. Then, we populate the dark-matter-only MDPL2 simulation with baryon properties and determine whether we can also successfully use the trained machine learning model on dark-matter-only low resolution simulations.

\subsection{Error analysis} 
In order to determine the accuracy of our ML models, we have trained our four models on the dataset composed of all features and on the dataset with the reduced set of features summarised in \autoref{table:important_features} using the experimental setup described in the previous section. The average performance of the models is shown in \autoref{fig:rMSE}. In the top panel, we show the MSE defined in Eq.(\ref{eq:lossf}) for the different tested models as a function of the target variables when all input features are used. In the bottom panel, the same quantities are displayed for the reduced set of features. 

As a general result, it can be observed from \autoref{fig:rMSE} that 
XGBoost algorithm has the best performance for all targets. For RF, we find equivalent performances for both sets of features in $\Mgas$, $\YX$ and $\YSZ$; a somewhat worse result for the reduced set  on $\Tgas$; and better performance on $\Mstar$ for the reduced set. For XGBoost, the trends  are similar to those of Random Forest, although the difference in performance for XGBoost between both sets of features is negligible for  $\Tgas$ and smaller for $\Mstar$. For the MLP model, all results using the reduced set of features are worse than those obtained when using all the features in  the catalogues. These differences between the tree based approaches (RF and XGBoost) and MLP can be explained taking into consideration that the selection of important features was  done  using Random Forest.  In any case, the performance of MLP is the worst for all targets even when all features are considered. For NGBoost, the MSE of mean predictions is similar in both set of features and worse than XGBoost and RF. After the previous analysis, we can conclude that XGBoost gives  the most accurate model predictions. However, as shown later, NGBoost is more accurate than XGBoost when it comes at mimicking the scatter of the true The300 data. Therefore, we will only consider XGBoost and NGBoost algorithms for the rest this work. A summary of the performance for all models can be found in \autoref{table:performance} for the reduced set of features.

\begin{figure*}
\includegraphics[width=0.97\textwidth]{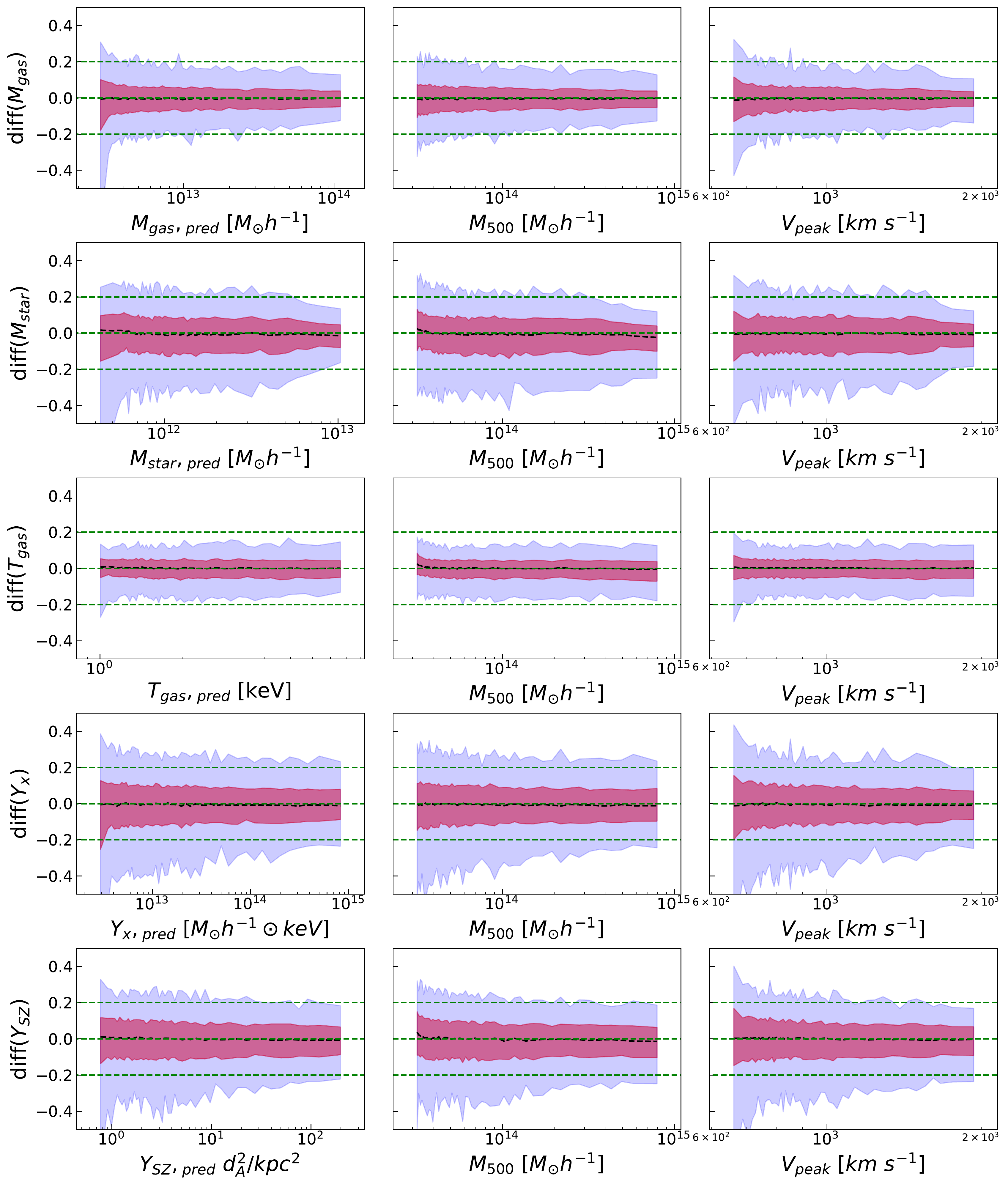}
\caption{The relative difference (y-axis) defined in Eq.(\ref{eq:error}) as a function of the XGBoost-predicted target variable (first column), the cluster 3D dynamical mass $M_{500c}(2)$ (second column) and the peak of the maximum value of the radial circular velocity profile  across the halo's  mass accretion history $V_{\text{peak}}(7)$ (third column). From top to bottom, each row corresponds to different baryon targets: $\Mgas$, $\Mstar$, $\Tgas$, $\YX$, $\YSZ$. Dashed black lines are the mean value of the relative difference, which are very close to diff=0. Red regions and blue regions represent the $1\sigma$ and $2\sigma$ scatter regions, respectively. Additionally, 0\% and 20\% relative difference lines are represented in dashed green colour. The data is binned using sliding windows that contain roughly the same number of clusters.}
\label{fig:errors}
\end{figure*}

 The scores shown in \autoref{fig:rMSE} summarise in a single value the performance of the models. However, they do not allow us to understand how the model performs in the different regions of the space of features and targets. In order to analyse this, we first define the relative difference in performance for a single target $\mathbf{y}$ as
\begin{equation}\label{eq:error}
    \text{diff}(\mathbf{y})=\frac{10^{<\log y_{\text{pred},i}|x_{i}>}-y_{\text{true}}}{10^{<\log y_{\text{pred},i}|x_{i}>}},
\end{equation}
 Note that in Eq.(\ref{eq:error}), we are not considering the logarithmic value of the targets, but the model aims at predicting the logarithmic values. One can interpret these differences as a probability distribution. This means that given a value of $y_{\text{pred}}$ one might estimate the aleatory scatter to that particular predicted value. These differences are shown in \autoref{fig:errors} as a function of the predicted target $y_{\text{pred}}$ (first column), the cluster mass $M_{500}$ (second column) and the peak of the velocity profile along the mass accretion history,  $V_{\text{peak}}$ (third column) for all redshifts. In \autoref{fig:errors}, instead of plotting the individual differences for all instances, the mean value (dashed black) and the 66\% (red region) and 95\% (blue region) confident intervals are represented for sliding windows (bins) containing roughly the same number of objects. 

The main result that can be observed from \autoref{fig:errors} is that 
the predictions are unbiased with respect to the most important features ($M_{500}$ and $V_{\text{peak}}$) and with respect to the predicted targets, since  the mean is very close to 0 for all ranges. However, the scatter varies depending on the target as it is depicted in \autoref{fig:rMSE} and \autoref{fig:errors}. Particularly, $T_{\text{gas}}$ is the target most accurately predicted, with an average scatter of 7\% (standard deviation of Eq.(\ref{eq:error})) and $Y_{\text{X}}$ is the  predicted variable with higher average scatter (~16\%).  The numerical values corresponding to 1 $\sigma$  of the distribution of the relative differences can be found in \autoref{table:performance}. In addition, we found a slight dependence of the scatter on the $M_{500}$, $V_{\text{peak}}$ and on the inferred targets values (except for $T_{\text{gas}}$). The scatter seems to decrease as these values increase. From a statistical point of view, the  scatter of baryon properties for high mass halos is smaller compared to low mass halos. A possible physical explanation is that massive clusters are more self-similar than smaller groups that present a larger halo-to-halo variation due to the stronger impact of non-gravitational processes. The relative difference for the  NGBoost mean predictions is similar to those of the other tested ML models (see \autoref{fig:errors}). However, overall scatter is higher, as shown in \autoref{table:performance}.

\begin{table*}
\centering
\caption{The MSE defined in Eq.(\ref{eq:lossf}) for the reduced set of features in \autoref{table:important_features}. In brackets, we show the standard deviation (scatter) $\sigma$ of the relative difference defined in Eq.(\ref{eq:error}). Rows correspond to values of the MSE for different models while columns correspond to values for different baryonic targets.}
\label{table:performance}
\begin{tabular}{cccccc} 
\toprule
 MSE $\times 10^{-3}$ & $\Mgas$     & $\Mstar$    & $\Tgas$     & $\YX$       & $\YSZ$       \\ 
\hline
XGBoost                  & 2.17 (11\%) & 3.43 (14\%) & 0.94 (7\%)  & 4.81 (17\%) & 4.07 (16\%)  \\
NGBoost                  & 2.58 (12\%) & 5.69 (18\%) & 1.45 (9\%)  & 5.71(19\%)  & 5.34 (19\%)  \\
RF                       & 2.20 (11\%) & 3.47 (14\%) & 1.56 (10\%) & 4.85 (17\%) & 3.96 (16\%)  \\
MLP                      & 2.75 (12\%) & 8.16 (21\%) & 2.23 (11\%) & 5.97 (19\%) & 5.82 (19\%)  \\
\bottomrule
\end{tabular}
\end{table*}

\subsection{XGBoost vs NGBoost}
\label{sec:NGBvsXGB}
As stated in section \autoref{sec-3}, NGBoost is a probabilistic model that can learn to infer not only the mean predictions, but also the scatter. In \autoref{table:performance}, we show that the best model is XGBoost only when taking into account its mean predicted values. However, when it comes at mimicking the complete behaviour of the data, the scatter of the predicted targets tends to be underestimated for deterministic ML models. 

\begin{figure}
\includegraphics[width=0.44\textwidth]{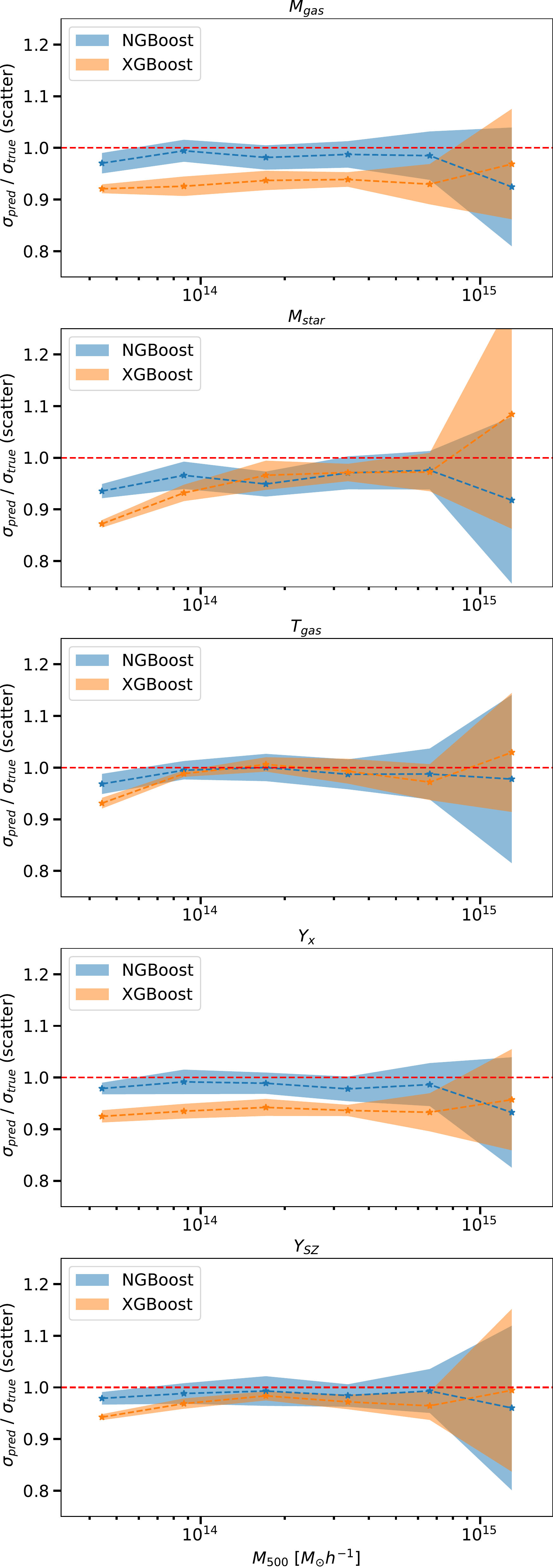} 
\caption{Predicted baryons' scatter divided by the true The300 scatter for different mass bins $M_{500}$ (x-axis). Dashed lines and shaded regions represent the mean and standard deviation for 10 different K-folds corresponding to NGBoost (blue) and XGBoost (orange) . From top to bottom,  the results are shown for different targets.}
\label{fig:NGBvsXGB}
\end{figure}

In order to test whether XGBoost and NGBoost predictions are similarly spread as in true The300 data, we bin our baryonic targets in mass $(M_{500})$ bins. This is done for The300 true targets, XGBoost and NGBoost predictions. Then, for every mass bin, the scatter (standard deviation) of the baryonic properties is computed. Note that this process is repeated ten times for our ten disjoint K-folds. The results are displayed in \autoref{fig:NGBvsXGB} where we show the standard deviation per mass bin of XGBoost and NGBoost predictions divided by the standard deviation of the true baryonic data. As a general result, NGBoost successfully mimics the true scatter for most targets. In contrast, XGBoost baryonic properties are $\sim 0.1 \text{dex}$ less spread. Notwithstanding, for $\Tgas$ and  $\YSZ$ the scatter of XGBoost is closer to the The300 true data with $\sigma_{\text{pred}}/\sigma_{\text{true}}=0.95$ on average. Also note that the scatter for $\Mstar$ is not completely well predicted by either of the models. However, NGBoost prediction of the scatter is also more  precise in this target. The reader should bear in mind that although NGBoost successfully mimics the behaviour of the data in terms of predicted scatter, the predicted baryonic values of the XGBoost models are always closer to The300 true data. 

In addition, we have also computed the covariance between different baryonic properties and checked that XGBoost and NGBoost predictions have a similar covariance structure than that of The300 simulation. The interested reader can find these results in the Appendix \ref{appendixB}.

\subsection{ML inference of Baryonic properties in Dark matter only datasets}
 We now proceed to apply the trained ML model to infer the different baryonic properties in the full set of MDPL2 halo catalogues.  We will use the 10 different XGBoost models trained  on the reduced set of features of The300 clusters.
In order to create the catalogue,  we first build a dataset with the reduced set of features (shown in \autoref{table:important_features}) for each halo of the full MDPL2 box. 
Note that the same transformations and cutoffs are applied to the full MDPL2 
Rockstar catalogue as in  \autoref{sec-2}. Next, we discard  clusters whose features values are not inside the hyper-cube defined by the MDPL2 features used for training since ML models are not designed for extrapolation inference. This means that only MDPL2 clusters such that
\begin{equation}
   x_{\text{min}}^{\text{training}}\leq x^{\text{MLDP2}}\leq x_{\text{max}}^{\text{training}} \text{ , for } x\in \text{features}
\end{equation}
 will be taken into consideration. Where $x^{\text{training}}$ is a feature corresponding to the  training dataset and $x^{\text{MLDP2}}$ is the same feature for the full MDPL2 simulation. Only 397 clusters out of 1,306,185 are outside the hyper-cube defined by the most important features and therefore, they are not considered for the analysis.   
 
\begin{figure*}
\includegraphics[width=0.9\textwidth]{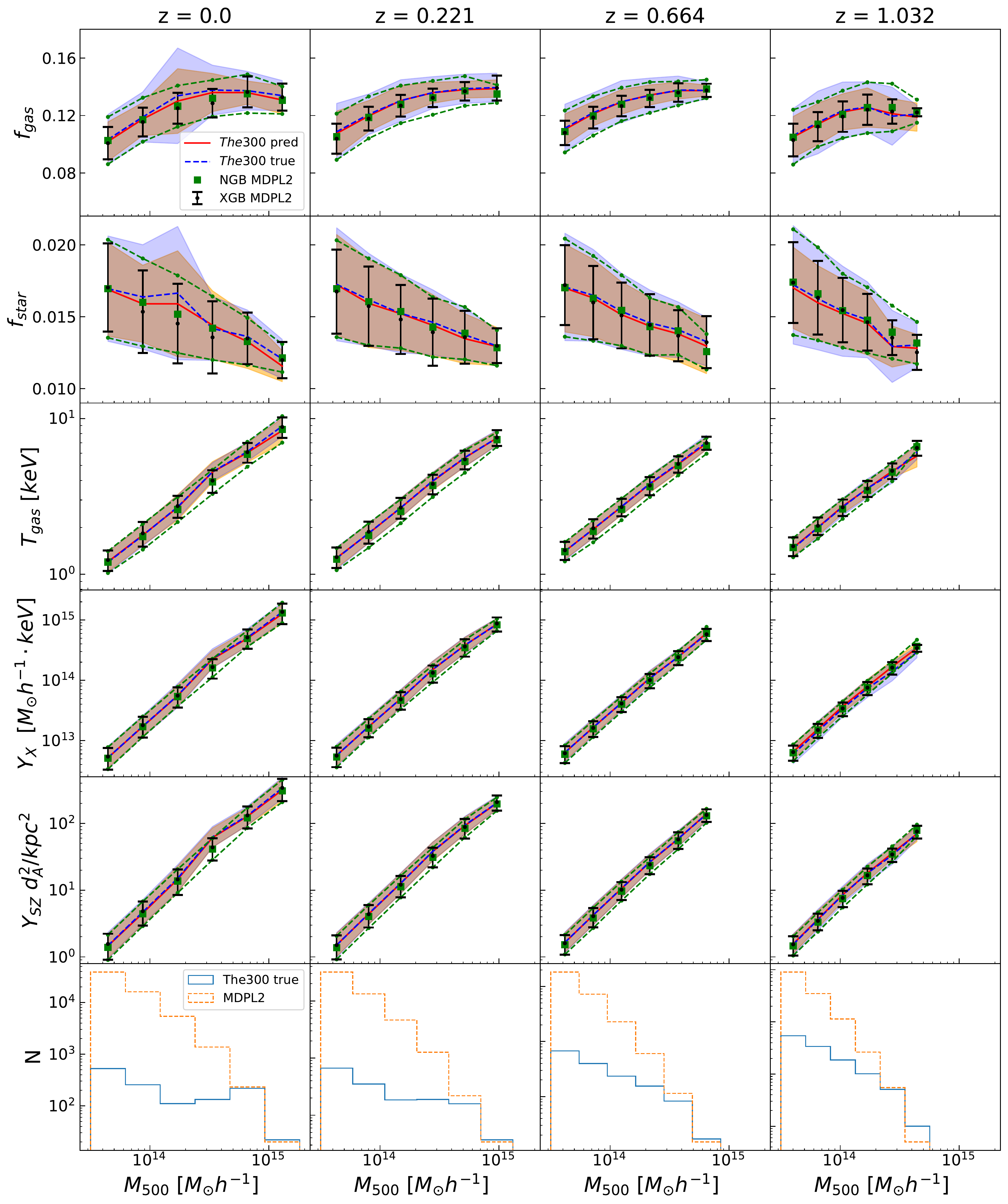}
\caption{XGBoost predictions (y-axis) as a function of cluster mass,  $M_{500c}(2)$, at  4 different redshifts (different columns) from $z=0$ to $z=1.032$. The first five rows correspond to the predictions of our baryonic targets: gas and star fractions $f_{\text{gas}}$ and $f_{\text{star}}$, gas temperature $\Tgas$, X-ray Y-parameter $\YX$ and SZ Y-parameter $\YSZ$. The data is binned along the x-axis and the means of the predicted values for the test set are shown in red dashed lines  with their scatter (standard deviation) represented as  shaded brown/orange region. True values of The300 train set are shown as a blue dashed line  and their scatter corresponds to shaded blue region. Black points represent the average values for the predictions of MDPL2 clusters per mass bin and the error bars correspond to  $1 \sigma$ scatter. The bottom row  shows  the number of cluster objects (N) per mass bin for The300 (blue histogram)  and MDPL2 (orange) simulations. In addition, we show in green squares the predictions corresponding to NGBoost for the full MDPL2 box.}
\label{fig:MDPL2}
\end{figure*}

 In order to evaluate if the generated catalogue presents properties that are coherent with the properties of fully simulated data, we will compare our baryon properties with the halo mass for The300 and the full MDPL2 generated catalogue. These results are shown 
in \autoref{fig:MDPL2} for different  redshift values (columns). In these plots, the values of the targets (rows)  are plotted with respect to  $M_{500}$. For the targets $\Mgas$ and $\Mstar$, the plots show  the relative fractions:
\begin{equation}
    f_{i}=M_{i}/M_{500}\text{ ,}
\end{equation}
where $i$ can be either gas or star. Error bars represent the intrinsic scatter of our baryon ML estimates on the MDPL2 catalogue ($1\sigma$), orange/brown regions correspond to $1\sigma$ region for The300  test set predictions (for all the k-folds) and blue regions are the equivalent but for The300 true targets. Moreover, in the last row of the figure the number of clusters per bin is represented as a function of $M_{500}$ for both The300 and MDPL2 datasets. 

As a general result, the XGBoost-predicted values for MDPL2  objects (black error bars) are similar and also their distributions per mass bin are comparable with the true values (blue region), i.e. in agreement with  \autoref{fig:errors}. However, the scatter of the predictions is slightly smaller (around 10-20\%) than the corresponding scatter using the true values of The300 data for $f_{\text{gas}}$ and $f_{\text{star}}$. This issue can be solved by using probabilistic regression models such us NGBoost where the scatter is  predicted more accurately. The mean predictions (green squares) and scatter (green dashed lines) for NGBoost is also shown in the same figure. Furthermore, a similar result to the ones shown in \autoref{fig:MDPL2} are obtained when plotting as a function of $\vpeak$ instead of $M_{500}$. We need to point out that for massive clusters ( $> 8\times 10^{14} \hMsun$), the number of objects is similar in the The300 and MDPL2 simulations. Particularly, the last two mass bins are mostly composed of the same objects and the difference lies in the baryon properties of the The300 simulation.

\subsubsection{Dependence of  ML model  predictions on DM mass resolution }
\label{sec:resolution}

The ML models have been trained on a particular DM simulation with a fixed resolution in mass. Here we are interested to compare the predictions of the ML model when applied to halo catalogues from  simulations with lower mass resolution. Since some of the features of the halos are expected to be affected by resolution, then the infer baryon quantities from the ML models could also be affect by that. Since our goal is to make our ML models as universal as possible so they can be applied to different DM-only simulations  with larger volumes, it is important to test for these effects. In order to do that, we are going to apply the trained XGBoost and NGBoost models in two simulations run with identical initial conditions but with a difference of a factor 8 in  particle mass. For this test, we are going to use also another completely different realisation  than  MDPL2, i.e. the  UNIT project.
The UNIT\footnote{\url{https://unitsims.ft.uam.es}} N-body cosmological simulations \citep[UNITSIM,][]{UNITSIM} are designed to provide accurate predictions of the  clustering properties of dark matter halos using the suppressed variance method proposed by \citet{suppress_variance}. We particularly focus on one of the  UNIT  simulations with the same box side length than  MDPL2, (i.e. $ 1 \hGpc$ ) and similar number of particles ($4096^3$). 
Furthermore, this  simulation has also  been performed with 8 times less number of particles  ($2048^{3}$). For simplicity we will  refer  to these two simulations as UNITSIM4096 and UNITSIM2048  for the  high and low resolution versions respectively.  

\begin{figure*}
\includegraphics[width=0.9\textwidth]{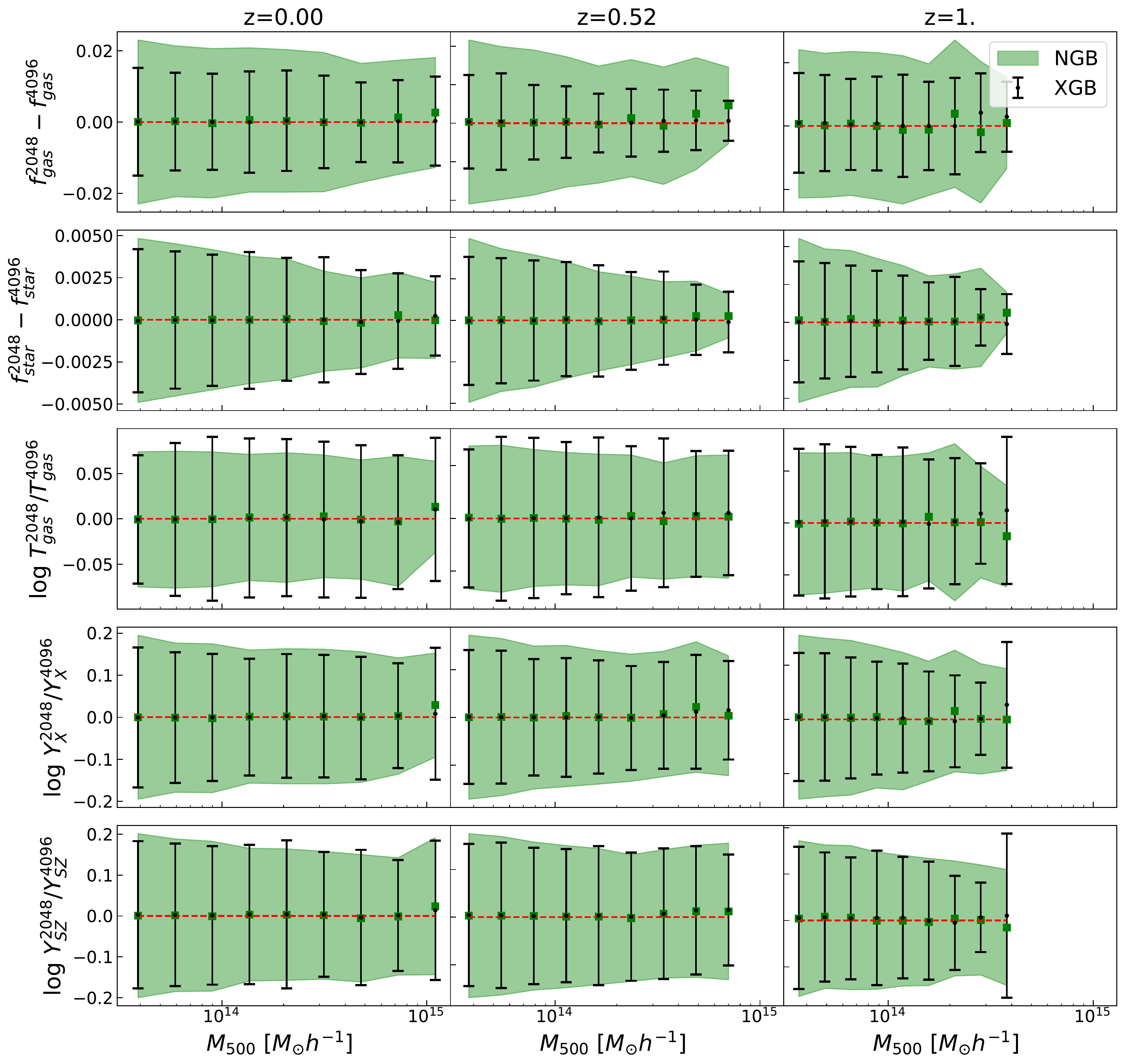}
\caption{The difference of XGBoost (error bars) and NGBoost (green shaded region) baryonic properties  for halos corresponding to the UNITSIM2048 and UNITSIM4096 DM-only simulations  as a function of the cluster total mass $M_{500c}(2)$. The y-axis represents values of $\bar{\mu}$ (Black points and green squares) and $\bar{\sigma}$ (error bars and green region) defined in Eq.(\ref{eq:barmu}) for different mass bins. From top to bottom, different rows represent the baryonic properties considered: $f_{\text{gas}}$, $f_{\text{star}}$, $\Tgas$, $\YX$ and $\YSZ$. From left to right, we show our results for three different redshifts z=0 (first column), z=0.52 (second column) and z=1 (third column). }

\label{fig:res}
\end{figure*}

Dark matter cluster-size halo catalogues  from {\sc Rockstar+ Consistent Trees} are then selected  for UNITSIM4096 and UNITSIM2048 following the same procedure described in \autoref{sec-2}. 
We then apply the trained XGBoost and NGBoost models to these catalogues to infer the target  baryon properties for each DM halos in the two versions. These baryon properties present similar statistics (mean and scatter per mass bin) as those  shown in \autoref{fig:MDPL2}. In order to make a more quantitative comparison of the results for the two UNIT simulations,  we bin the data as in \autoref{fig:MDPL2} according to  $M_{500}$ and compute the difference of the mean values and estimate an upper limit for its  scatter as
\begin{equation}\label{eq:barmu}
    \bar{\mu} = \mu_{2048}-\mu_{4096} \hspace{2mm}\text{and}\hspace{2mm} \bar{\sigma}=\sqrt{\sigma_{2048}^{2}+\sigma_{4096}^{2}}.
\end{equation}
Here, $\mu$ stands for mean values and $\sigma$ for the standard deviation of a bin.  The particular values of $\bar{\mu}$ and $\bar{\sigma}$ are shown for 3 different snapshots in \autoref{fig:res}. 
As can be seen in this  figure, $\bar{\mu}\simeq0$ with a small value for  the scatter  for all mass  bins.  The scatter $\bar{\sigma}$ is within $\sim 2\%$ for $f_{\text{gas}}$ and $\sim 0.5\%$ for $f_{\text{star}}$. For $\Tgas$,  the residuals amount to $\sim 0.1$ dex and for $\YX$ and $\YSZ$ up to $\sim 0.2$ dex. Therefore, we conclude that the baryonic properties predicted by the ML model for the same halos simulated  with a factor of 8 difference in mass resolution are statistically equivalent. NGBoost predictions have slightly larger scatter overall (green shaded area) and the mean values (green squares) are similar to XGBoost predictions.

\section{Validation of the Gas Scaling Relations}\label{sec-5}
 Scaling relations are generally power laws that relate properties in astrophysical systems, such us the Colour-Magnitude Relation or the  Tully Fisher relation \citep{tully1977new} for galaxies.  The applications of the scaling  relations are manifold, such as inferring masses of galaxy clusters that are sensitive to cosmological parameters \citep[e.g.][]{CosmoPlanck}. For a recent review of scaling relations for  galaxy clusters we refer the reader to e.g. \cite{lovisari2022scalingrelations}.  The temperature-mass relation can be written as 
 \begin{equation}\label{eq:srT}
 E(z)^{-2/3} \frac{\Tgas}{\text{keV}}  = 10^{\AT}\left(\frac{M}{\Msun}\right)^{\BT}\text{, }
 \end{equation}
 where $E(z) = H(z)/H_{0}$ and H(z) is the Hubble parameter. Similarly, for the $\YX-M$ and $\YSZ-M$ we use
 \begin{equation}\label{eq:srX}
    E(z)^{-2/3}  \frac{\YX}{\hMsun\text{keV}} =  10^{\AX}\left(\frac{M}{\Msun}\right)^{\BX}
 \end{equation}
 and
 \begin{equation}\label{eq:srSZ}
     E(z)^{-2/3}  \frac{d_{A}^{2}\YSZ}{\Mpc^{2}} =  10^{\ASZ}\left(\frac{M}{\Msun}\right)^{\BSZ}\text{.}
 \end{equation}
Here, $A_{i}$ and $B_{i}$ ($i=T,\text{X},\text{SZ}$) are the parameters that we are interested in obtaining by fitting the above equations to our data.  Once we have generated baryon catalogues for different N-body simulations we apply a simple linear fitting function in logarithm space to fit the data to the equations listed above. However, selecting data from different snapshots gives us small variations of the $A_{i}$ and $B_{i}$  best fitting parameters with redshift. Therefore, we use the following parametrization to study the redshift dependence:
 \begin{equation}\label{eq:srupdateA}
     A_{i}(z) = A_{i,0}(1+z)^{\alpha_{i}}
 \end{equation}
 \begin{equation}\label{eq:updateB}
     B_{i}(z) = B_{i,0}(1+z)^{\beta_{i}}
 \end{equation}
 where $A_{i,0}$ and $B_{i,0}$ are the values of the intercept and slope at z=0 and $\alpha_{i}$ and $\beta_{i}$ describe their possible dependence with respect to the redshift. With this new parametrization we apply a non-linear least square fitting model to fit the function described by equations (\ref{eq:srT}), (\ref{eq:srX}) and (\ref{eq:srSZ}) updated with equations (\ref{eq:srupdateA}) and (\ref{eq:updateB}). The best-fitting parameters are shown in Tables \ref{table:SL1}, \ref{table:SLredshift1}, \ref{table:SL} and \ref{table:SLredshift}.  Note that we have used the mass corresponding to the N-body simulation (the feature variable $M_{500c}(2)$ as the mass of the cluster). Moreover, in order to study the dependence of these parameters on the mass, we opt for a similar approach as \cite{brokenlaw}, i.e. a fixed broken power law. We split the data in different mass ranges and fit the above equations. The considered mass ranges are: 1) $\log M [\hMsun]>13.5$ (all our data), 2) $\log M [\hMsun]>14$, 3) $\log M [\hMsun]\leq 14$. We have considered this particular pivot point $\log M [\hMsun]=14$ because all of the scaling relations appear to break at that point for the radiative models.  Furthermore, in order to study how the data is spread around the scaling law, we also used the MSE defined in Eq.(\ref{eq:lossf}) and the relative error defined in Eq.(\ref{eq:error}). Note that here the  mean predictions $<y_{\text{pred}}| x >$ will be given by the corresponding scaling law. 
 
 As a general result, the fitting parameters are in agreement among the three different N-body simulations and are slightly  different from The300 hydrodynamical simulation. This deviation, though small, is caused by the fact of considering the full box of dark matter only simulations instead of the smaller volume of the `zoom' simulation. The effect of resolution is negligible for galaxy clusters. There is also a small difference between The300 simulations true data (The300), and the fitting counterpart using the ML predicted data (The300*). This slight difference can be mainly appreciated in the intrinsic scatter of the linear fitting function, which is generally smaller in the case of The300*. It is important to note that the scatter of the scaling law for The300 simulations is generally larger when comparing it with the values shown in \autoref{table:performance}, where the scatter (standard deviation of the relative difference) is reduced by a factor of 0.5 for the gas temperature, 0.3 for $\YX$ and 0.45 for $\YSZ$. 
Moreover, the most relevant variables for each gas properties presented in table \autoref{table:important_features} can be used for finding analytical expressions for scaling laws with a reduced MSE using genetic algorithms \citep{CAMELSSZ}. The difference between NGBoost and XGBoost is that the MSE with respect to the scaling law is generally bigger in the case of NGBoost-predicted baryonic properties. This means that probabilistic regression mimics the scatter of the true data more accurately than non-probabilistic models.

 As far as the redshift dependence is concerned, it is negligible for $\YX$ and $\YSZ$ where the parameters $\alpha$ and $\beta$ are of order $\lesssim10^{-3}$. However, the parameter $\alpha_{T}\simeq 0.3$ cannot be ignored. This indicates that the evolution of $\Tgas$ is relevant as it can also be appreciated in \autoref{table:important_features}, where the scale factor a(24) is the second most important variable, reducing the normalised MSE from 1 to 0.6. Regarding the mass dependence, the scaling law parameters can vary around  $\sim10\%$, for instance the slope of the $\YSZ$ can be 1.615 ($\log M [\hMsun]>14$) or 1.8 ($\log M [\hMsun]\leq14$) when using The300 data. An interesting result is that the mass dependence is less prominent when we consider the full statistics of the whole MDPL2 simulation, in which the slope of the $\YSZ$ can vary from  1.677 ($\log M [\hMsun]>14$) or 1.730 ($\log M [\hMsun]\leq14$).

\begin{table*}
\centering
\caption{\textbf{XGBoost best fit parameters:} The best fit  parameters for the $T-M$, $\YX-M$ and $\YSZ-M$ relations for the different simulation sets. The MSE in Eq.(\ref{eq:lossf}) and average scatter of the relative difference in Eq.(\ref{eq:error}), (in parenthesis), are  also shown. For The300 simulation, the true values of the baryon properties have been used while for The300* the predicted XGBoost values are used instead. The relative error in the estimated parameters $A_{i}$, and $B_{i}$ is always $\leq 10^{-3}$.}
\label{table:SL1}
\begin{tabular}{cccccccccc} 
\toprule
\begin{tabular}[c]{@{}c@{}}\\\\Simulation\end{tabular} & ${\AT}_{,0}$ & ${\BT}_{,0}$ & $\text{MSE}_T$     & ${\AX}_{,0}$ & ${\BX}_{,0}$ & $\text{MSE}_\text{X}$ & ${\ASZ}_{,0}$ & ${\BSZ}_{,0}$ & $\text{ MSE}_\text{SZ}$  \\ 
\hline
\multicolumn{10}{c}{$\log M [\hMsun] > 13.5$}                                                                                                                                                                  \\
The300                                                 & 0.2083   & 0.6081   & $1.8\times10^{-3}$(10\%) & 13.09    & ~1.718   & $8.3\times10^{-3}$(25\%)    & -5.499    & 1.697     & $9.5\times10^{-3}$(23\%)      \\
The300*                                                & 0.2082   & 0.6054   & $1.6\times10^{-3}$(10\%) & 13.08    & ~1.718   & $3.9\times10^{-3}$(19\%)    & -5.497    & 1.692     & $7.2\times10^{-3}$(20\%)      \\
MDPL2                                                  & 0.2133   & 0.5863   & $3.3\times10^{-3}$(11\%) & 13.07    & 1.767    & $2.8\times10^{-3}$(13\%)    & -5.513    & 1.710     & $6.5\times10^{-3}$(21\%)      \\
UNITSIM4096                                            & 0.2122   & 0.5865   & $3.3\times10^{-3}$(11\%) & 13.07    & 1.767    & $2.8\times10^{-3}$(13\%)    & -5.514    & 1.709     & $6.5\times10^{-3}$(21\%)      \\
UNITSIM2048                                            & 0.2126   & 0.5854   & $3.3\times10^{-3}$(11\%) & 13.07    & 1.766    & $2.8\times10^{-3}$(13\%)    & -5.515    & 1.709     & $6.5\times10^{-3}$(21\%)      \\ 
\hline
\multicolumn{10}{c}{$\log M [\hMsun]>14$}                                                                                                                                                                      \\
The300                                                 & 0.2121   & 0.6081   & $1.4\times10^{-3}$(8\%)  & 13.14    & 1.642    & $5.7\times10^{-3}$(19\%)    & -5.436    & 1.615     & $6.1\times10^{-3}$(19\%)      \\
The300*                                                & 0.2117   & 0.6023   & $1.4\times10^{-3}$(8\%)  & 13.14    & 1.643    & $2.5\times10^{-3}$(11\%)    & -5.438    & 1.614     & $5.0\times10^{-3}$(16\%)      \\
MDPL2                                                  & 0.2197   & 0.5829   & $2.4\times10^{-3}$(11\%) & 13.10    & 1.673    & $1.8\times10^{-3}$(10\%)    & -5.501    & 1.665     & $5.1\times10^{-3}$(17\%)      \\
UNITSIM4096                                            & 0.2202   & 0.5803   & $2.4\times10^{-3}$(11\%) & 13.10    & 1.668    & $1.8\times10^{-3}$(11\%)    & -5.502    & 1.664     & $5.4\times10^{-3}$(22\%)      \\
UNITSIM2048                                            & 0.2200   & 0.5803   & $2.4\times10^{-3}$(11\%) & 13.10    & 1.668    & $1.9\times10^{-3}$(12\%)    & -5.501    & 1.664     & $5.5\times10^{-3}$(22\%)      \\ 
\hline
\multicolumn{10}{c}{$\log M [\hMsun]\leq 14$}                                                                                                                                                                                                                                                                           \\
The300                                                 & 0.2052   & 0.5928   & $2.0\times10^{-3}$(10\%) & 13.09    & 1.827    & $9.3\times10^{-3}$(26\%)    & -5.490    & 1.800     & $11\times10^{-3}$(25\%)       \\
The300*                                                & 0.2030   & 0.5746   & $1.6\times10^{-3}$(9\%)  & 13.09    & 1.825    & $4.3\times10^{-3}$(15\%)    & -5.491    & 1.767     & $8.0\times10^{-3}$(20\%)      \\
MDPL2                                                  & 0.2090   & 0.5632   & $2.1\times10^{-3}$(10\%) & 13.08    & 1.827    & $2.9\times10^{-3}$(13\%)    & -5.505    & 1.753     & $6.6\times10^{-3}$(19\%)      \\
UNITSIM4096                                            & 0.2076   & 0.5622   & $2.1\times10^{-3}$(10\%) & 13.08    & 1.829    & $2.9\times10^{-3}$(13\%)    & -5.506    & 1.751     & $6.6\times10^{-3}$(19\%)      \\
UNITSIM2048                                            & 0.2081   & 0.5622   & $2.1\times10^{-3}$(10\%) & 13.08    & 1.828    & $2.9\times10^{-3}$(13\%)    & -5.508    & 1.753     & $6.6\times10^{-3}$(19\%)      \\
\bottomrule
\end{tabular}
\end{table*}

\begin{table*}
\centering
\caption{\textbf{XGBoost best fit redshift dependence parameters} for the scaling relations defines in \autoref{eq:srupdateA} and \autoref{eq:updateB} }
\label{table:SLredshift1}
\begin{tabular}{ccccccc} 
\toprule
\begin{tabular}[c]{@{}c@{}}\\Simulation\end{tabular} & $\alpha_T(\times 10^{-3})$ & $\beta_T(\times10^{-3})$ & $\alpha_\text{X}(\times10^{-3})$ & $\beta_\text{X}(\times10^{-3})$ & $\alpha_\text{SZ} (\times10^{-3}) $ & $\beta_\text{SZ} (\times10^{-3})$  \\ 
\hline
The300                                               & $-339.0\pm 5.3$          & $-3.8\pm4.3$           & $-1.15\pm0.16$                 & $3.4\pm3.3$                   & $5.36\pm0.41$                     & $-11.3\pm 3.5$                \\
The300*                                              & $-336.7\pm 4.8$          & $-11.1\pm4.0$          & $-0.73\pm0.11$                 & $6.4\pm2.2$                   & $6.22\pm0.45$                     & $-12.2\pm3.1$                 \\
MDPL2                                                & $-314.3\pm 1.2$          & $30.4\pm1.8$           & $0.417\pm0.021$                & $0.42\pm0.65$                 & $7.181\pm0.074$                   & $0.2\pm 1.0$                     \\
UNITSIM4096                                          & $-308.1\pm 1.3$          & $30.5\pm1.8$           & $0.52\pm0.21$                  & $0.22\pm0.68$                 & $7.162\pm0.075$                   & $0.0\pm 1.1$                     \\
UNITSIM2048                                          & $-302.9\pm 1.3$          & $29.9\pm1.8$           & $-0.052\pm0.021$               & $-0.93\pm0.71$                & $6.279\pm0.075$                   & $-2.0\pm1.1$                     \\ 
\hline
\multicolumn{7}{c}{$\log M [\hMsun]>14$}                                                                                                                                                                                                         \\
The300                                               & $-332\pm 18$             & $-20\pm10$             & $-5.28\pm0.52$                 & $11.4\pm7.4$                  & $17.1\pm1.3$                      & $3.7\pm7.7$                      \\
The300*                                              & $-329\pm 18$             & $-25\pm10$             & $-4.96\pm0.35$                 & $16.8\pm4.9$                  & $17.4\pm1.2$                      & $2.9\pm 7.0$                     \\
MDPL2                                                & $-375.3\pm 8.2$          & $40.2\pm 7.2$          & $-2.17\pm0.10$                 & $17.9\pm2.2$                  & $7.23\pm0.42$                     & $-1.1\pm 3.8$                    \\
UNITSIM4096                                          & $-382.0\pm 9.0$          & $42.3\pm 7.9$          & $-2.06\pm0.11$                 & $18.9\pm2.4$                  & $6.60\pm0.47$                     & $-2.1\pm 4.2$                    \\
UNITSIM2048                                          & $-387.3\pm 9.0$          & $53.2\pm 8.0$          & $-2.03\pm0.11$                 & $20.2\pm2.4$                  & $6.13\pm0.47$                     & $-3.2\pm 4.2$                    \\ 
\hline
\multicolumn{7}{c}{$\log M [\hMsun]\leq14$}                                                                                                                                                                                                      \\
The300                                               & $-301.4\pm 9.0$          & $69\pm14$              & $-1.39\pm0.27$                 & $-41\pm10$                    & $5.37\pm0.70$                     & $-48\pm11$                       \\
The300*                                              & $-294.7\pm 8.1$          & $81\pm13$              & $-0.89\pm0.18$                 & $-35.2\pm7.1$                 & $5.52\pm0.60$                     & $-31\pm10$                       \\
MDPL2                                                & $-269.9\pm 1.8$          & $115.4\pm2.9$          & $-1.011\pm0.031$               & $-32.0\pm1.1$                 & $8.87\pm0.11$                     & $-33.6\pm1.7$                    \\
UNITSIM4096                                          & $-259.8\pm 1.8$          & $121.9\pm2.8$          & $-1.063\pm0.031$               & $-35.2\pm1.1$                 & $9.01\pm0.11$                     & $-37.5\pm1.7$                    \\
UNITSIM2048                                          & $-251.9\pm 1.8$          & $122.8\pm2.9$          & $-0.698\pm0.031$               & $-33.7\pm1.2$                 & $7.96\pm 0.11$                    & $-35.2\pm1.8$                    \\
\bottomrule
\end{tabular}
\end{table*}

\begin{table*}
\centering
\caption{\textbf{NGBoost best fit parameters:} The best fit  parameters for the $T-M$, $\YX-M$ and $\YSZ-M$ relations for the different simulation sets. The MSE in Eq.(\ref{eq:lossf}) and average scatter of the relative difference in Eq.(\ref{eq:error}), (in parenthesis), are  also shown. For The300 simulation, the true values of the baryon properties have been used while for The300* the predicted NGBoost values are used instead. The relative error in the estimated parameters $A_{i}$, and $B_{i}$ is always $\leq 10^{-3}$.}
\label{table:SL}
\begin{tabular}{cccccccccc} 
\toprule
\begin{tabular}[c]{@{}c@{}}\\\\Simulation\end{tabular} & ${\AT}_{,0}$ & ${\BT}_{,0}$ & $\text{MSE}_T$     & ${\AX}_{,0}$ & ${\BX}_{,0}$ & $\text{MSE}_\text{X}$ & ${\ASZ}_{,0}$ & ${\BSZ}_{,0}$ & $\text{MSE}_\text{SZ}$  \\ 
\hline
\multicolumn{10}{c}{$\log M [\hMsun] > 13.5$}                                                                                                                                                                 \\
The300                                                 & 0.2083   & 0.6081   & $1.8\times10^{-3}$(10\%) & 13.09    & ~1.718   & $8.3\times10^{-3}$(24\%)    & -5.499    & 1.697     & $9.5\times10^{-3}$(23\%)      \\
The300*                                                & 0.2085   & 0.6089   & $1.6\times10^{-3}$(10\%) & 13.08    & ~1.715   & $7.0\times10^{-3}$(20\%)    & -5.499    & 1.697     & $8.3\times10^{-3}$(21\%)      \\
MDPL2                                                  & 0.2000   & 0.5726   & $3.5\times10^{-3}$(13\%) & 13.07    & 1.768    & $7.2\times10^{-3}$(21\%)    & -5.525    & 1.716     & $11\times10^{-3}$(27\%)       \\
UNITSIM4096                                            & 0.2000   & 0.5716   & $3.5\times10^{-3}$(13\%) & 13.07    & 1.770    & $7.3\times10^{-3}$22\%)     & -5.524    & 1.712     & $11\times10^{-3}$(27\%)       \\
UNITSIM2048                                            & 0.1989   & 0.5726   & $3.5\times10^{-3}$(14\%) & 13.07    & 1.769    & $7.3\times10^{-3}$(23\%)    & -5.526    & 1.712     & $12\times10^{-3}$(31\%)       \\ 
\hline
\multicolumn{10}{c}{$\log M [\hMsun]>14$}                                                                                                                                                                     \\
The300                                                 & 0.2121   & 0.6054   & $1.4\times10^{-3}$(8\%)  & 13.14    & 1.642    & $5.7\times10^{-3}$(19\%)    & -5.436    & 1.615     & $6.1\times10^{-3}$(19\%)      \\
The300*                                                & 0.2142   & 0.6025   & $1.4\times10^{-3}$(8\%)  & 13.14    & 1.646    & $4.9\times10^{-3}$(16\%)    & -5.430    & 1.610     & $5.5\times10^{-3}$(17\%)      \\
MDPL2                                                  & 0.1895   & 0.6155   & $3.7\times10^{-3}$(14\%) & 13.10    & 1.688    & $4.6\times10^{-3}$(16\%)    & -5.510    & 1.677     & $9.5\times10^{-3}$(23\%)      \\
UNITSIM4096                                            & 0.1890   & 0.6149   & $3.7\times10^{-3}$(14\%) & 13.10    & 1.660    & $4.9\times10^{-3}$(20\%)    & -5.513    & 1.674     & $9.9\times10^{-3}$(29\%)      \\
UNITSIM2048                                            & 0.1890   & 0.6147   & $3.7\times10^{-3}$(14\%) & 13.10    & 1.675    & $5.0\times10^{-3}$(21\%)    & -5.510    & 1.675     & $9.8\times10^{-3}$(28\%)      \\ 
\hline
\multicolumn{10}{c}{$\log M [\hMsun]\leq14$}                                                                                                                                                                     \\
The300                                                 & 0.2052   & 0.5928   & $2.0\times10^{-3}$(10\%) & 13.09    & 1.827    & $9.3\times10^{-3}$(25\%)    & -5.490    & 1.800     & $11\times10^{-3}$(25\%)       \\
The300*                                                & 0.2061   & 0.5908   & $1.8\times10^{-3}$(10\%) & 13.09    & 1.820    & $8.0\times10^{-3}$(23\%)    & -5.491    & 1.794     & $9.5\times10^{-3}$(23\%)      \\
MDPL2                                                  & 0.1915   & 0.5300   & $3.4\times10^{-3}$(13\%) & 13.08    & 1.824    & $7.6\times10^{-3}$(21\%)    & -5.523    & 1.732     & $12\times10^{-3}$(27\%)       \\
UNITSIM4096                                            & 0.1917   & 0.5287   & $3.4\times10^{-3}$(13\%) & 13.08    & 1.828    & $7.7\times10^{-3}$(22\%)    & -5.523    & 1.732     & $12\times10^{-3}$(27\%)       \\
UNITSIM2048                                            & 0.1914   & 0.5289   & $3.4\times10^{-3}$(13\%) & 13.08    & 1.828    & $7.6\times10^{-3}$(22\%)    & -5.521    & 1.730     & $12\times10^{-3}$(27\%)       \\
\bottomrule
\end{tabular}
\end{table*}

\begin{table*}
\centering
\caption{\textbf{NGBoost best fit redshift dependence parameters} for the scaling relations defines in \autoref{eq:srupdateA} and \autoref{eq:updateB} }
\label{table:SLredshift}
\begin{tabular}{ccccccc} 
\toprule
\begin{tabular}[c]{@{}c@{}}\\Simulation\end{tabular} & $\alpha_T(\times 10^{-3})$ & $\beta_T(\times10^{-3})$ & $\alpha_\text{X}(\times10^{-3})$ & $\beta_\text{X}(\times10^{-3})$ & $\alpha_\text{SZ} (\times10^{-3}) $ & $\beta_\text{SZ} (\times10^{-3})$  \\ 
\hline
The300                                               & $-339.0\pm 5.3$          & $-3.8\pm4.3$           & $-1.15\pm0.16$                 & $3.4\pm3.3$                   & $5.36\pm0.41$                     & $-11.3\pm 3.5$                   \\
The300*                                              & $-345.8\pm 4.9$          & $-4.1\pm4.1$           & $-0.49\pm0.15$                 & $11.2\pm3.9$                  & $5.69\pm0.38$                     & $-10.3\pm3.3$                    \\
MDPL2                                                & $-308.8\pm 1.6$          & $14.5\pm2.3$           & $-0.565\pm0.033$               & $-4.2\pm1.0$                  & $3.809\pm0.099$                   & $-13.2\pm 1.4$                   \\
UNITSIM4096                                          & $-322.6\pm 1.7$          & $20.1\pm2.2$           & $-0.628\pm0.032$               & $-2.2\pm1.1$                  & $5.449\pm0.097$                   & $-5.6\pm 1.4$                    \\
UNITSIM2048                                          & $-313.5\pm 1.7$          & $20.0\pm2.3$           & $-0.465\pm0.033$               & $-5.7\pm1.7$                  & $4.580\pm0.095$                   & $-6.8\pm1.4$                     \\ 
\hline
\multicolumn{7}{c}{$\log M [\hMsun]>14$}                                                                                                                                                                                                         \\
The300                                               & $-332\pm 18$             & $-20\pm10$             & $-5.28\pm0.52$                 & $11.4\pm7.4$                  & $17.1\pm1.3$                      & $3.7\pm7.7$                      \\
The300*                                              & $-369\pm 18$             & $-7\pm10$              & $-3.89\pm0.47$                 & $4.8\pm6.6$                   & $18.1\pm1.2$                      & $1.7\pm 7.3$                     \\
MDPL2                                                & $-276\pm 11$             & $-19.9\pm 8.8$         & $-1.57\pm0.17$                 & $-2.4\pm3.5$                  & $5.06\pm0.58$                     & $-9.3\pm 7.3$                    \\
UNITSIM4096                                          & $-289.0\pm 12$           & $-18.6\pm 9.3$         & $-1.84\pm0.18$                 & $1.30\pm3.8$                  & $5.94\pm0.61$                     & $-7.6\pm 5.5$                    \\
UNITSIM2048                                          & $-290\pm 14$             & $-21.0\pm 9.4$         & $-1.70\pm0.17$                 & $1.25\pm3.9$                  & $5.51\pm0.61$                     & $-11.3\pm 5.5$                   \\ 
\hline
\multicolumn{7}{c}{$\log M [\hMsun]\leq14$}                                                                                                                                                                                                      \\
The300                                               & $-301.4\pm 9.0$          & $69\pm14$              & $-1.39\pm0.27$                 & $-41\pm10$                    & $-5.37\pm0.70$                    & $-48\pm11$                       \\
The300*                                              & $-314.3\pm 8.3$          & $44\pm13$              & $-1.07\pm0.25$                 & $-33\pm10$                    & $6.34\pm0.65$                     & $-37\pm11$                       \\
MDPL2                                                & $-266.9\pm 2.5$          & $113.5\pm3.8$          & $-0.986\pm0.049$               & $-31.1\pm1.7$                 & $4.13\pm0.14$                     & $-21.2\pm2.3$                    \\
UNITSIM4096                                          & $-275.8\pm 2.4$          & $132.8\pm3.6$          & $-1.244\pm0.048$               & $-33.5\pm1.7$                 & $6.22\pm0.15$                     & $-17.3\pm2.2$                    \\
UNITSIM2048                                          & $-264.1\pm 2.5$          & $128.7\pm3.7$          & $-0.973\pm0.048$               & $-33.7\pm1.7$                 & $4.85\pm 0.14$                    & $-22.0\pm2.2$                    \\
\bottomrule
\end{tabular}
\end{table*}

\section{SUMMARY AND CONCLUSIONS}\label{sec-6}
Numerical simulations are key to studying galaxy clusters. On the one hand, with the current technology it is possible to perform large volume N-body simulations  that can be useful to describe the dark-matter component. However, big volume hydrodynamical simulations cannot be carried out due to their computational demands. We have therefore trained a set of machine learning models to populate high volume dark-matter-only simulations with baryonic properties. In particular, we have defined our feature space as the {\sc Rockstar} variables of DM-only halos and our target variables are directly estimated from \thethreehundred{} hydrodynamical simulations: the mass of the gas $\Mgas$, the mass of the stars $\Mstar$, the gas temperature $\Tgas$, X-ray Y-parameter $\YX$ and the integrated Compton-y parameter $\YSZ$. All these quantities are integrated quantities in spherical region of overdensity 500 times the critical density at their corresponding redshift.

Particularly, we have considered four different ML models, random forest (RF), extreme gradient boosting (XGBoost), MultiLayer Perceptron (MLP) and Natural Gradient Boosting for Probabilistic Prediction (NGBoost). We have determined that XGBoost is the algorithm that is more suitable to our dataset and whose predictions are closer to the true hydrodynamical targets, as shown in \autoref{table:performance}. However, as depicted in \autoref{fig:NGBvsXGB}, probabilistic regression is needed for a proper modelling of the scatter and therefore, NGBoost is the best model in that regard. We have applied an algorithm --\textit{Greedy Search Feature Importance Algorithm (GSFIA)}-- to identify the features that have more predictive information. By using GSFIA, we have managed to reduce the dimensionality of our feature space from 27 to approximately 5 variables depending on the target variable. We have demonstrated that masses and velocities have a higher amount of predictive information while time evolution variables play a secondary role in the prediction of our targets. What is more, ellipticity, dynamical state, and spin features are redundant. A possible explanation for this is that our baryon targets are integrated in spherical regions. 

Then, we have applied our trained ML models to populate halo catalogues  with baryonic properties from two full box N-body simulations: the {\sc MultiDark} simulation (MDPL2) and the UNIT N-body cosmological simulations (UNITSIM). The MDPL2 predicted baryon properties are compatible to those of The300 simulations, as shown  in \autoref{fig:MDPL2}. The application on two UNITSIM simulations with $1\hGpc$ box size $2048^{3}$ and $4046^{3}$ particles has determined that our model can be successfully applied to boxes whose resolution is up to $1/8$ of the corresponding simulation used for training. This suggests that this is a promising method to populate the  UNITSIM large volume N-body  halos  with baryon properties up to $27 \Gpc^{3}$ (i.e a $3 \hGpc$ box size  with $6144^3$ particles). This  will be an excellent tool to  study the large scale distribution of  galaxy clusters in an unprecedented way. For instance,  we can estimate  the cosmic variance in the number counts of X-ray detected clusters from  the  eROSITA satellite  all-sky survey \cite{liu2021erosita} by extracting many different light-cones  from  this large computational volume. This will be the subject  of a forthcoming paper. 

Furthermore, the scaling relations are powerful mass-observable proxies. We have check that the best-fitting parameters inferred using our three mock DM full-box baryon catalogues are compatible. They nevertheless differ slightly from those of the The300, partially because of  the considerable smaller number of cluster objects in the hydrodynamical  simulations  used to get the best fit values. This would suggest  that  mass  completeness  have an small impact, thought not negligible, in the calibration of the mass-proxies. 

Our  ML models have been trained using the results from a simulation  with fixed cosmological parameters and a particular selection of values for the parameters of the subgrid physics models implemented in GADGETX  \citep{cui2018three}.  This is clearly a limitation of the applicability of the ML models  presented in this paper for simulations with different cosmologies. This limitation could be alleviated  by running an ensemble of thousands of simulations varying cosmological and astrophysical parameters  and training the  ML models so they can  marginalise over all the parameters (either explicitly as feature variables or implicitly inside other features such as mass, etc). A recent example of this method is the CAMELS project 
\citep{CAMELS}.  Unfortunately, given the low density of cluster sized objects,  the simulations needed to  repeat the CAMELS technique for galaxy clusters would require of  Gigaparsec volumes with  multi-billion dark matter particles and then, using  the zooming technique,  generating many hydrodynamical simulations  of the same region with different baryon physics models. This is certainly a way to go for the future, but it is well beyond the scope of this paper.

Moreover, our test concerning the dependence of ML models predictions on DM mass resolution in \autoref{sec:resolution}  suggests that the joint distribution of halo properties ($M$, $V_{\text{peak}}$, etc) is not very sensitive of the numerical resolution.

To conclude, our  work  shows that ML models are very useful methods for finding a mapping between dark matter halo properties found in N-body and the complex  hydrodynamical simulations. We checked that, on average, the generated catalogue for the 3 dark-matter-only simulations used throughout this paper have the same distributions to that of true training set and therefore, they can be used for painting dark matter halos  with baryonic properties that are directly related with observed quantities, providing added value to large volume collisionless N-body simulations.

\section*{Acknowledgements}
The authors thank the anonymous referee for his/her invaluable comments and suggestions, without which this work would be incomplete. D.d.A., W.C. and G.Y.  would like to thank Ministerio de Ciencia e Innovación for financial support under project grant PID2021-122603NB-C21.
WC is supported by the STFC AGP Grant ST/V000594/1 and the Atracci\'{o}n de Talento Contract no. 2020-T1/TIC-19882 granted by the Comunidad de Madrid in Spain. He also thanks the Ministerio de Ciencia e Innovación (Spain) for financial support under Project grant PID2021-122603NB-C21. He further acknowledges the science research grants from the China Manned Space Project with NO. CMS-CSST-2021-A01 and CMS-CSST-2021-B01.
G.M. acknowledges financial support from PID2019-106827GB-I00/AEI / 10.13039/501100011033
The {\sc CosmoSim} database used in this paper is a service by the Leibniz-Institute for Astrophysics Potsdam (AIP).
The {\sc MultiDark} database was developed in cooperation with the Spanish MultiDark Consolider Project CSD2009-00064.
The authors acknowledge The Red Española de Supercomputación for
granting  computing time for running  the hydrodynamical simulations of The300 galaxy cluster  project in  the Marenostrum supercomputer at the
Barcelona Supercomputing Center.

\section*{DATA AVAILABILITY}
The trained models and data products for MDPL2, UNITSIM2048 and UNITSIM4096 are publicly available at \url{https://github.com/The300th/DarkML}.

\vspace{3mm}

\newcommand{\newblock}{}
\bibliographystyle{mnras}
\bibliography{refs}

\begin{thebibliography}{}
\makeatletter
\relax
\def\mn@urlcharsother{\let\do\@makeother \do\$\do\&\do\#\do\^\do\_\do\%\do\~}
\def\mn@doi{\begingroup\mn@urlcharsother \@ifnextchar [ {\mn@doi@}
  {\mn@doi@[]}}
\def\mn@doi@[#1]#2{\def\@tempa{#1}\ifx\@tempa\@empty \href
  {http://dx.doi.org/#2} {doi:#2}\else \href {http://dx.doi.org/#2} {#1}\fi
  \endgroup}
\def\mn@eprint#1#2{\mn@eprint@#1:#2::\@nil}
\def\mn@eprint@arXiv#1{\href {http://arxiv.org/abs/#1} {{\tt arXiv:#1}}}
\def\mn@eprint@dblp#1{\href {http://dblp.uni-trier.de/rec/bibtex/#1.xml}
  {dblp:#1}}
\def\mn@eprint@#1:#2:#3:#4\@nil{\def\@tempa {#1}\def\@tempb {#2}\def\@tempc
  {#3}\ifx \@tempc \@empty \let \@tempc \@tempb \let \@tempb \@tempa \fi \ifx
  \@tempb \@empty \def\@tempb {arXiv}\fi \@ifundefined
  {mn@eprint@\@tempb}{\@tempb:\@tempc}{\expandafter \expandafter \csname
  mn@eprint@\@tempb\endcsname \expandafter{\@tempc}}}

\bibitem[\protect\citeauthoryear{Allen, Evrard  \& Mantz}{Allen
  et~al.}{2011}]{allen2011cosmological}
Allen S.~W.,  Evrard A.~E.,   Mantz A.~B.,  2011, Annual Review of Astronomy
  and Astrophysics, 49, 409

\bibitem[\protect\citeauthoryear{{Allgood}, {Flores}, {Primack}, {Kravtsov},
  {Wechsler}, {Faltenbacher}  \& {Bullock}}{{Allgood}
  et~al.}{2006}]{allgood2006}
{Allgood} B.,  {Flores} R.~A.,  {Primack} J.~R.,  {Kravtsov} A.~V.,  {Wechsler}
  R.~H.,  {Faltenbacher} A.,   {Bullock} J.~S.,  2006, \mn@doi [\mnras]
  {10.1111/j.1365-2966.2006.10094.x}, \href
  {https://ui.adsabs.harvard.edu/abs/2006MNRAS.367.1781A} {367, 1781}

\bibitem[\protect\citeauthoryear{Altmann, Toloşi, Sander  \& Lengauer}{Altmann
  et~al.}{2010}]{Altmann20101340}
Altmann A.,  Toloşi L.,  Sander O.,   Lengauer T.,  2010, Bioinformatics, 26,
  1340

\bibitem[\protect\citeauthoryear{{Angulo} \& {Pontzen}}{{Angulo} \&
  {Pontzen}}{2016}]{suppress_variance}
{Angulo} R.~E.,  {Pontzen} A.,  2016, \mn@doi [\mnras] {10.1093/mnrasl/slw098},
  \href {https://ui.adsabs.harvard.edu/abs/2016MNRAS.462L...1A} {462, L1}

\bibitem[\protect\citeauthoryear{Angulo, Springel, White, Jenkins, Baugh  \&
  Frenk}{Angulo et~al.}{2012}]{Langulo2012MillenniumXX}
Angulo R.,  Springel V.,  White S.,  Jenkins A.,  Baugh C.,   Frenk C.,  2012,
  Monthly Notices of the Royal Astronomical Society, 426, 2046

\bibitem[\protect\citeauthoryear{{Angulo}, {Zennaro}, {Contreras}, {Aric{\`o}},
  {Pellejero-Iba{\~n}ez}  \& {St{\"u}cker}}{{Angulo}
  et~al.}{2021}]{2021MNRAS.507.5869A}
{Angulo} R.~E.,  {Zennaro} M.,  {Contreras} S.,  {Aric{\`o}} G.,
  {Pellejero-Iba{\~n}ez} M.,   {St{\"u}cker} J.,  2021, \mn@doi [\mnras]
  {10.1093/mnras/stab2018}, \href
  {https://ui.adsabs.harvard.edu/abs/2021MNRAS.507.5869A} {507, 5869}

\bibitem[\protect\citeauthoryear{{Arnaud}, {Pratt}, {Piffaretti},
  {B{\"o}hringer}, {Croston}  \& {Pointecouteau}}{{Arnaud}
  et~al.}{2010}]{Arnaud2010}
{Arnaud} M.,  {Pratt} G.~W.,  {Piffaretti} R.,  {B{\"o}hringer} H.,  {Croston}
  J.~H.,   {Pointecouteau} E.,  2010, \mn@doi [\aap]
  {10.1051/0004-6361/200913416}, \href
  {https://ui.adsabs.harvard.edu/abs/2010A&A...517A..92A} {517, A92}

\bibitem[\protect\citeauthoryear{Bah{\'e} et~al.,}{Bah{\'e}
  et~al.}{2017}]{bahe2017hydrangea}
Bah{\'e} Y.~M.,  et~al., 2017, Monthly Notices of the Royal Astronomical
  Society, 470, 4186

\bibitem[\protect\citeauthoryear{Barnes, Kay, Henson, McCarthy, Schaye  \&
  Jenkins}{Barnes et~al.}{2016}]{barnes2016macsis}
Barnes D.~J.,  Kay S.~T.,  Henson M.~A.,  McCarthy I.~G.,  Schaye J.,   Jenkins
  A.,  2016, Monthly Notices of the Royal Astronomical Society, p. stw2722

\bibitem[\protect\citeauthoryear{Barnes et~al.,}{Barnes
  et~al.}{2017}]{barnes2017eagle}
Barnes D.~J.,  et~al., 2017, Monthly Notices of the Royal Astronomical Society,
  471, 1088

\bibitem[\protect\citeauthoryear{{Barredo Arrieta} et~al.,}{{Barredo Arrieta}
  et~al.}{2020}]{BARREDOARRIETA202082}
{Barredo Arrieta} A.,  et~al., 2020, \mn@doi [Information Fusion]
  {https://doi.org/10.1016/j.inffus.2019.12.012}, 58, 82

\bibitem[\protect\citeauthoryear{{Baugh}}{{Baugh}}{2006}]{2006RPPh...69.3101B}
{Baugh} C.~M.,  2006, \mn@doi [Reports on Progress in Physics]
  {10.1088/0034-4885/69/12/R02}, \href
  {https://ui.adsabs.harvard.edu/abs/2006RPPh...69.3101B} {69, 3101}

\bibitem[\protect\citeauthoryear{Behroozi, Wechsler  \& Wu}{Behroozi
  et~al.}{2012}]{Rockstar}
Behroozi P.~S.,  Wechsler R.~H.,   Wu H.-Y.,  2012, \mn@doi [The Astrophysical
  Journal] {10.1088/0004-637x/762/2/109}, 762, 109

\bibitem[\protect\citeauthoryear{{Behroozi}, {Wechsler}, {Wu}, {Busha},
  {Klypin}  \& {Primack}}{{Behroozi} et~al.}{2013}]{2013ApJ...763...18B}
{Behroozi} P.~S.,  {Wechsler} R.~H.,  {Wu} H.-Y.,  {Busha} M.~T.,  {Klypin}
  A.~A.,   {Primack} J.~R.,  2013, \mn@doi [\apj] {10.1088/0004-637X/763/1/18},
  \href {https://ui.adsabs.harvard.edu/abs/2013ApJ...763...18B} {763, 18}

\bibitem[\protect\citeauthoryear{Benson}{Benson}{2012}]{benson2012galacticus}
Benson A.~J.,  2012, New Astronomy, 17, 175

\bibitem[\protect\citeauthoryear{Bent\'ejac, Cs\"org\H{o}  \&
  {Martínez-Mu\~noz}}{Bent\'ejac et~al.}{2021}]{bentejac21comp}
Bent\'ejac C.,  Cs\"org\H{o} A.,   {Martínez-Mu\~noz} G.,  2021, Artificial
  Intelligence Review, 54, 1937

\bibitem[\protect\citeauthoryear{{Bernardini}, {Feldmann},
  {Angl{\'e}s-Alc{\'a}zar}, {Boylan-Kolchin}, {Bullock}, {Mayer}  \&
  {Stadel}}{{Bernardini} et~al.}{2022}]{bernardinibaryonfields}
{Bernardini} M.,  {Feldmann} R.,  {Angl{\'e}s-Alc{\'a}zar} D.,
  {Boylan-Kolchin} M.,  {Bullock} J.,  {Mayer} L.,   {Stadel} J.,  2022,
  \mn@doi [\mnras] {10.1093/mnras/stab3088}, \href
  {https://ui.adsabs.harvard.edu/abs/2022MNRAS.509.1323B} {509, 1323}

\bibitem[\protect\citeauthoryear{{Borgani} et~al.,}{{Borgani}
  et~al.}{2004}]{Borgani2004}
{Borgani} S.,  et~al., 2004, \mn@doi [\mnras]
  {10.1111/j.1365-2966.2004.07431.x}, \href
  {https://ui.adsabs.harvard.edu/abs/2004MNRAS.348.1078B} {348, 1078}

\bibitem[\protect\citeauthoryear{Breiman}{Breiman}{2001}]{breiman01rf}
Breiman L.,  2001, Machine Learning, 45, 5

\bibitem[\protect\citeauthoryear{{Bryan} \& {Norman}}{{Bryan} \&
  {Norman}}{1998}]{bryan199820years}
{Bryan} G.~L.,  {Norman} M.~L.,  1998, \mn@doi [\apj] {10.1086/305262}, \href
  {https://ui.adsabs.harvard.edu/abs/1998ApJ...495...80B} {495, 80}

\bibitem[\protect\citeauthoryear{{Bullock}, {Kolatt}, {Sigad}, {Somerville},
  {Kravtsov}, {Klypin}, {Primack}  \& {Dekel}}{{Bullock}
  et~al.}{2001}]{bullockspin}
{Bullock} J.~S.,  {Kolatt} T.~S.,  {Sigad} Y.,  {Somerville} R.~S.,  {Kravtsov}
  A.~V.,  {Klypin} A.~A.,  {Primack} J.~R.,   {Dekel} A.,  2001, \mn@doi
  [\mnras] {10.1046/j.1365-8711.2001.04068.x}, \href
  {https://ui.adsabs.harvard.edu/abs/2001MNRAS.321..559B} {321, 559}

\bibitem[\protect\citeauthoryear{Chen \& Guestrin}{Chen \&
  Guestrin}{2016}]{chen16XGBoost}
Chen T.,  Guestrin C.,  2016, in Proceedings of the 22Nd ACM SIGKDD
  International Conference on Knowledge Discovery and Data Mining. KDD '16.
ACM, New York, NY, USA, pp 785--794

\bibitem[\protect\citeauthoryear{{Chisari} et~al.,}{{Chisari}
  et~al.}{2016}]{2016MNRAS.461.2702C}
{Chisari} N.,  et~al., 2016, \mn@doi [\mnras] {10.1093/mnras/stw1409}, \href
  {https://ui.adsabs.harvard.edu/abs/2016MNRAS.461.2702C} {461, 2702}

\bibitem[\protect\citeauthoryear{{Chuang} et~al.,}{{Chuang}
  et~al.}{2019}]{UNITSIM}
{Chuang} C.-H.,  et~al., 2019, \mn@doi [\mnras] {10.1093/mnras/stz1233}, \href
  {https://ui.adsabs.harvard.edu/abs/2019MNRAS.487...48C} {487, 48}

\bibitem[\protect\citeauthoryear{{Cora} et~al.,}{{Cora}
  et~al.}{2018}]{cora2018SAG}
{Cora} S.~A.,  et~al., 2018, \mn@doi [\mnras] {10.1093/mnras/sty1131}, \href
  {https://ui.adsabs.harvard.edu/\#abs/2018MNRAS.479....2C} {479, 2}

\bibitem[\protect\citeauthoryear{{Croton} et~al.,}{{Croton}
  et~al.}{2016}]{croton2016SAGE}
{Croton} D.~J.,  et~al., 2016, \mn@doi [\apjs] {10.3847/0067-0049/222/2/22},
  \href {https://ui.adsabs.harvard.edu/\#abs/2016ApJS..222...22C} {222, 22}

\bibitem[\protect\citeauthoryear{{Cui}, {Borgani}, {Dolag}, {Murante}  \&
  {Tornatore}}{{Cui} et~al.}{2012}]{Cui2012}
{Cui} W.,  {Borgani} S.,  {Dolag} K.,  {Murante} G.,   {Tornatore} L.,  2012,
  \mn@doi [\mnras] {10.1111/j.1365-2966.2012.21037.x}, \href
  {https://ui.adsabs.harvard.edu/abs/2012MNRAS.423.2279C} {423, 2279}

\bibitem[\protect\citeauthoryear{{Cui}, {Borgani}  \& {Murante}}{{Cui}
  et~al.}{2014}]{Cui2014}
{Cui} W.,  {Borgani} S.,   {Murante} G.,  2014, \mn@doi [\mnras]
  {10.1093/mnras/stu673}, \href
  {https://ui.adsabs.harvard.edu/abs/2014MNRAS.441.1769C} {441, 1769}

\bibitem[\protect\citeauthoryear{Cui et~al.,}{Cui et~al.}{2018}]{cui2018three}
Cui W.,  et~al., 2018, Monthly Notices of the Royal Astronomical Society, 480,
  2898

\bibitem[\protect\citeauthoryear{{Cui} et~al.,}{{Cui} et~al.}{2022}]{Cui2022}
{Cui} W.,  et~al., 2022, \mn@doi [\mnras] {10.1093/mnras/stac1402}, \href
  {https://ui.adsabs.harvard.edu/abs/2022MNRAS.514..977C} {514, 977}

\bibitem[\protect\citeauthoryear{{\VAN{Daniel}{de Andres}{de Andres}}
  et~al.,}{{\VAN{Daniel}{de Andres}{de Andres}}
  et~al.}{2022}]{deAndres2022Planck}
{\VAN{Daniel}{de Andres}{de Andres}} D.,  et~al., 2022, \mn@doi [Nature
  Astronomy] {10.1038/s41550-022-01784-y}, \href
  {https://ui.adsabs.harvard.edu/abs/2022NatAs.tmp..218D} {}

\bibitem[\protect\citeauthoryear{{Dav{\'e}}, {Angl{\'e}s-Alc{\'a}zar},
  {Narayanan}, {Li}, {Rafieferantsoa}  \& {Appleby}}{{Dav{\'e}}
  et~al.}{2019}]{Dave2019}
{Dav{\'e}} R.,  {Angl{\'e}s-Alc{\'a}zar} D.,  {Narayanan} D.,  {Li} Q.,
  {Rafieferantsoa} M.~H.,   {Appleby} S.,  2019, \mn@doi [\mnras]
  {10.1093/mnras/stz937}, \href
  {https://ui.adsabs.harvard.edu/abs/2019MNRAS.486.2827D} {486, 2827}

\bibitem[\protect\citeauthoryear{Dietterich}{Dietterich}{1998}]{dietterich98machinelearning}
Dietterich T.~G.,  1998, AI MAGAZINE, 18, 97

\bibitem[\protect\citeauthoryear{Dolag, Komatsu  \& Sunyaev}{Dolag
  et~al.}{2016}]{dolag2016magneticum}
Dolag K.,  Komatsu E.,   Sunyaev R.,  2016, Monthly Notices of the Royal
  Astronomical Society, 463, 1797

\bibitem[\protect\citeauthoryear{{Duan}, {Avati}, {Ding}, {Thai}, {Basu}, {Ng}
  \& {Schuler}}{{Duan} et~al.}{2019}]{NGBoost2019}
{Duan} T.,  {Avati} A.,  {Ding} D.~Y.,  {Thai} K.~K.,  {Basu} S.,  {Ng} A.~Y.,
   {Schuler} A.,  2019, arXiv e-prints, \href
  {https://ui.adsabs.harvard.edu/abs/2019arXiv191003225D} {p. arXiv:1910.03225}

\bibitem[\protect\citeauthoryear{{Eisert}, {Pillepich}, {Nelson}, {Klessen},
  {Huertas-Company}  \& {Rodriguez-Gomez}}{{Eisert}
  et~al.}{2022}]{Uncertainties4}
{Eisert} L.,  {Pillepich} A.,  {Nelson} D.,  {Klessen} R.~S.,
  {Huertas-Company} M.,   {Rodriguez-Gomez} V.,  2022, arXiv e-prints, \href
  {https://ui.adsabs.harvard.edu/abs/2022arXiv220206967E} {p. arXiv:2202.06967}

\bibitem[\protect\citeauthoryear{{Evrard}, {Metzler}  \& {Navarro}}{{Evrard}
  et~al.}{1996}]{evrard199620years}
{Evrard} A.~E.,  {Metzler} C.~A.,   {Navarro} J.~F.,  1996, \mn@doi [\apj]
  {10.1086/177798}, \href
  {https://ui.adsabs.harvard.edu/abs/1996ApJ...469..494E} {469, 494}

\bibitem[\protect\citeauthoryear{Fern\'andez-Delgado, Cernadas, Barro  \&
  Amorim}{Fern\'andez-Delgado et~al.}{2014}]{fernandez14comp}
Fern\'andez-Delgado M.,  Cernadas E.,  Barro S.,   Amorim D.,  2014, Journal of
  Machine Learning Research, 15, 3133

\bibitem[\protect\citeauthoryear{Ferri, Pudil, Hatef  \& Kittler}{Ferri
  et~al.}{1994}]{ferri1994comparative}
Ferri F.~J.,  Pudil P.,  Hatef M.,   Kittler J.,  1994, in , Vol.~16, Machine
  Intelligence and Pattern Recognition.
Elsevier, pp 403--413

\bibitem[\protect\citeauthoryear{Fosalba, Crocce, Gazta{\~n}aga  \&
  Castander}{Fosalba et~al.}{2015}]{fosalba2015mice}
Fosalba P.,  Crocce M.,  Gazta{\~n}aga E.,   Castander F.,  2015, Monthly
  Notices of the Royal Astronomical Society, 448, 2987

\bibitem[\protect\citeauthoryear{Habib et~al.,}{Habib
  et~al.}{2016}]{habib2016hacc}
Habib S.,  et~al., 2016, New Astronomy, 42, 49

\bibitem[\protect\citeauthoryear{{Ho}, {Farahi}, {Rau}  \& {Trac}}{{Ho}
  et~al.}{2021}]{Uncertainties3}
{Ho} M.,  {Farahi} A.,  {Rau} M.~M.,   {Trac} H.,  2021, \mn@doi [\apj]
  {10.3847/1538-4357/abd101}, \href
  {https://ui.adsabs.harvard.edu/abs/2021ApJ...908..204H} {908, 204}

\bibitem[\protect\citeauthoryear{{Ishiyama} et~al.,}{{Ishiyama}
  et~al.}{2021}]{2021MNRAS.506.4210I}
{Ishiyama} T.,  et~al., 2021, \mn@doi [\mnras] {10.1093/mnras/stab1755}, \href
  {https://ui.adsabs.harvard.edu/abs/2021MNRAS.506.4210I} {506, 4210}

\bibitem[\protect\citeauthoryear{Jo \& Kim}{Jo \& Kim}{2019}]{jo2019machine}
Jo Y.,  Kim J.-h.,  2019, Monthly Notices of the Royal Astronomical Society,
  489, 3565

\bibitem[\protect\citeauthoryear{{Kamdar}, {Turk}  \& {Brunner}}{{Kamdar}
  et~al.}{2016}]{Kamdar2016-zm}
{Kamdar} H.~M.,  {Turk} M.~J.,   {Brunner} R.~J.,  2016, \mn@doi [\mnras]
  {10.1093/mnras/stv2981}, \href
  {https://ui.adsabs.harvard.edu/\#abs/2016MNRAS.457.1162K} {457, 1162}

\bibitem[\protect\citeauthoryear{{Klypin}, {Trujillo-Gomez}  \&
  {Primack}}{{Klypin} et~al.}{2011}]{2011ApJ...740..102K}
{Klypin} A.~A.,  {Trujillo-Gomez} S.,   {Primack} J.,  2011, \mn@doi [\apj]
  {10.1088/0004-637X/740/2/102}, \href
  {https://ui.adsabs.harvard.edu/abs/2011ApJ...740..102K} {740, 102}

\bibitem[\protect\citeauthoryear{{Klypin}, {Yepes}, {Gottl{\"o}ber}, {Prada}
  \& {He{\ss}}}{{Klypin} et~al.}{2016}]{klypin2016multidark}
{Klypin} A.,  {Yepes} G.,  {Gottl{\"o}ber} S.,  {Prada} F.,   {He{\ss}} S.,
  2016, \mn@doi [\mnras] {10.1093/mnras/stw248}, \href
  {https://ui.adsabs.harvard.edu/\#abs/2016MNRAS.457.4340K} {457, 4340}

\bibitem[\protect\citeauthoryear{{Knollmann} \& {Knebe}}{{Knollmann} \&
  {Knebe}}{2009}]{AHF}
{Knollmann} S.~R.,  {Knebe} A.,  2009, \mn@doi [\apjs]
  {10.1088/0067-0049/182/2/608}, \href
  {https://ui.adsabs.harvard.edu/abs/2009ApJS..182..608K} {182, 608}

\bibitem[\protect\citeauthoryear{{Kodi Ramanah}, {Wojtak}, {Ansari}, {Gall}  \&
  {Hjorth}}{{Kodi Ramanah} et~al.}{2020}]{Uncertainties1}
{Kodi Ramanah} D.,  {Wojtak} R.,  {Ansari} Z.,  {Gall} C.,   {Hjorth} J.,
  2020, \mn@doi [\mnras] {10.1093/mnras/staa2886}, \href
  {https://ui.adsabs.harvard.edu/abs/2020MNRAS.499.1985K} {499, 1985}

\bibitem[\protect\citeauthoryear{{Kodi Ramanah}, {Wojtak}  \& {Arendse}}{{Kodi
  Ramanah} et~al.}{2021}]{Uncertainties2}
{Kodi Ramanah} D.,  {Wojtak} R.,   {Arendse} N.,  2021, \mn@doi [\mnras]
  {10.1093/mnras/staa3922}, \href
  {https://ui.adsabs.harvard.edu/abs/2021MNRAS.501.4080K} {501, 4080}

\bibitem[\protect\citeauthoryear{Kravtsov \& Borgani}{Kravtsov \&
  Borgani}{2012}]{kravtsov2012formation}
Kravtsov A.~V.,  Borgani S.,  2012, Annual Review of Astronomy and
  Astrophysics, 50, 353

\bibitem[\protect\citeauthoryear{{Kravtsov}, {Vikhlinin}  \&
  {Nagai}}{{Kravtsov} et~al.}{2006}]{2006ApJ...650..128K}
{Kravtsov} A.~V.,  {Vikhlinin} A.,   {Nagai} D.,  2006, \mn@doi [\apj]
  {10.1086/506319}, \href
  {https://ui.adsabs.harvard.edu/abs/2006ApJ...650..128K} {650, 128}

\bibitem[\protect\citeauthoryear{Kuhn, Johnson  et~al.}{Kuhn
  et~al.}{2013}]{kuhn2013applied}
Kuhn M.,  Johnson K.,   et~al., 2013, Applied predictive modeling.
~ Vol. 26, Springer

\bibitem[\protect\citeauthoryear{{Lacey} et~al.,}{{Lacey}
  et~al.}{2016}]{galform}
{Lacey} C.~G.,  et~al., 2016, \mn@doi [\mnras] {10.1093/mnras/stw1888}, \href
  {https://ui.adsabs.harvard.edu/abs/2016MNRAS.462.3854L} {462, 3854}

\bibitem[\protect\citeauthoryear{Le~Brun, McCarthy  \& Melin}{Le~Brun
  et~al.}{2015}]{le2015testing}
Le~Brun A.~M.,  McCarthy I.~G.,   Melin J.-B.,  2015, Monthly Notices of the
  Royal Astronomical Society, 451, 3868

\bibitem[\protect\citeauthoryear{{Le Brun}, {McCarthy}, {Schaye}  \&
  {Ponman}}{{Le Brun} et~al.}{2017}]{brokenlaw}
{Le Brun} A. M.~C.,  {McCarthy} I.~G.,  {Schaye} J.,   {Ponman} T.~J.,  2017,
  \mn@doi [\mnras] {10.1093/mnras/stw3361}, \href
  {https://ui.adsabs.harvard.edu/abs/2017MNRAS.466.4442L} {466, 4442}

\bibitem[\protect\citeauthoryear{{Li} et~al.,}{{Li} et~al.}{2020}]{Li2020}
{Li} Q.,  et~al., 2020, \mn@doi [\mnras] {10.1093/mnras/staa1385}, \href
  {https://ui.adsabs.harvard.edu/abs/2020MNRAS.495.2930L} {495, 2930}

\bibitem[\protect\citeauthoryear{Liu et~al.,}{Liu
  et~al.}{2021}]{liu2021erosita}
Liu A.,  et~al., 2021, arXiv preprint arXiv:2106.14518

\bibitem[\protect\citeauthoryear{{Lovell}, {Wilkins}, {Thomas}, {Schaller},
  {Baugh}, {Fabbian}  \& {Bah{\'e}}}{{Lovell} et~al.}{2022}]{LovellCeagleML}
{Lovell} C.~C.,  {Wilkins} S.~M.,  {Thomas} P.~A.,  {Schaller} M.,  {Baugh}
  C.~M.,  {Fabbian} G.,   {Bah{\'e}} Y.,  2022, \mn@doi [\mnras]
  {10.1093/mnras/stab3221}, \href
  {https://ui.adsabs.harvard.edu/abs/2022MNRAS.509.5046L} {509, 5046}

\bibitem[\protect\citeauthoryear{{Lovisari} \& {Maughan}}{{Lovisari} \&
  {Maughan}}{2022}]{lovisari2022scalingrelations}
{Lovisari} L.,  {Maughan} B.~J.,  2022, arXiv e-prints, \href
  {https://ui.adsabs.harvard.edu/abs/2022arXiv220207673L} {p. arXiv:2202.07673}

\bibitem[\protect\citeauthoryear{McCarthy, Bird, Schaye, Harnois-Deraps, Font
  \& Van~Waerbeke}{McCarthy et~al.}{2018}]{mccarthy2018bahamas}
McCarthy I.~G.,  Bird S.,  Schaye J.,  Harnois-Deraps J.,  Font A.~S.,
  Van~Waerbeke L.,  2018, Monthly Notices of the Royal Astronomical Society,
  476, 2999

\bibitem[\protect\citeauthoryear{{McGibbon} \& {Khochfar}}{{McGibbon} \&
  {Khochfar}}{2022}]{McGibbonMultiepoch}
{McGibbon} R.~J.,  {Khochfar} S.,  2022, \mn@doi [\mnras]
  {10.1093/mnras/stac1269}, \href
  {https://ui.adsabs.harvard.edu/abs/2022MNRAS.513.5423M} {513, 5423}

\bibitem[\protect\citeauthoryear{Moews, Dav{\'e}, Mitra, Hassan  \& Cui}{Moews
  et~al.}{2021}]{moews2021hybrid}
Moews B.,  Dav{\'e} R.,  Mitra S.,  Hassan S.,   Cui W.,  2021, Monthly Notices
  of the Royal Astronomical Society, 504, 4024

\bibitem[\protect\citeauthoryear{{Murante}, {Monaco}, {Giovalli}, {Borgani}  \&
  {Diaferio}}{{Murante} et~al.}{2010}]{Murante2010}
{Murante} G.,  {Monaco} P.,  {Giovalli} M.,  {Borgani} S.,   {Diaferio} A.,
  2010, \mn@doi [\mnras] {10.1111/j.1365-2966.2010.16567.x}, \href
  {https://ui.adsabs.harvard.edu/abs/2010MNRAS.405.1491M} {405, 1491}

\bibitem[\protect\citeauthoryear{{Navarro}, {Frenk}  \& {White}}{{Navarro}
  et~al.}{1997}]{1997ApJ...490..493N}
{Navarro} J.~F.,  {Frenk} C.~S.,   {White} S. D.~M.,  1997, \mn@doi [\apj]
  {10.1086/304888}, \href
  {https://ui.adsabs.harvard.edu/abs/1997ApJ...490..493N} {490, 493}

\bibitem[\protect\citeauthoryear{{Nelson} et~al.,}{{Nelson}
  et~al.}{2019}]{nelson2019illustris}
{Nelson} D.,  et~al., 2019, \mn@doi [Computational Astrophysics and Cosmology]
  {10.1186/s40668-019-0028-x}, \href
  {https://ui.adsabs.harvard.edu/abs/2019ComAC...6....2N} {6, 2}

\bibitem[\protect\citeauthoryear{Nembrini, K{\"o}nig  \& Wright}{Nembrini
  et~al.}{2018}]{nembrini2018revival}
Nembrini S.,  K{\"o}nig I.~R.,   Wright M.~N.,  2018, Bioinformatics, 34, 3711

\bibitem[\protect\citeauthoryear{{Osato} \& {Nagai}}{{Osato} \&
  {Nagai}}{2022}]{baryonpastingOsato}
{Osato} K.,  {Nagai} D.,  2022, arXiv e-prints, \href
  {https://ui.adsabs.harvard.edu/abs/2022arXiv220102632O} {p. arXiv:2201.02632}

\bibitem[\protect\citeauthoryear{Pedregosa et~al.,}{Pedregosa
  et~al.}{2011}]{scikit-learn}
Pedregosa F.,  et~al., 2011, Journal of Machine Learning Research, 12, 2825

\bibitem[\protect\citeauthoryear{{Peebles}}{{Peebles}}{1969}]{1969ApJ...155..393P}
{Peebles} P.~J.~E.,  1969, \mn@doi [\apj] {10.1086/149876}, \href
  {https://ui.adsabs.harvard.edu/abs/1969ApJ...155..393P} {155, 393}

\bibitem[\protect\citeauthoryear{{Planck Collaboration} et~al.,}{{Planck
  Collaboration} et~al.}{2016}]{CosmoPlanck}
{Planck Collaboration} et~al., 2016, \mn@doi [\aap]
  {10.1051/0004-6361/201525830}, \href
  {https://ui.adsabs.harvard.edu/abs/2016A&A...594A..13P} {594, A13}

\bibitem[\protect\citeauthoryear{Planelles, Borgani, Dolag, Ettori, Fabjan,
  Murante  \& Tornatore}{Planelles et~al.}{2013}]{planelles2013baryon}
Planelles S.,  Borgani S.,  Dolag K.,  Ettori S.,  Fabjan D.,  Murante G.,
  Tornatore L.,  2013, Monthly Notices of the Royal Astronomical Society, 431,
  1487

\bibitem[\protect\citeauthoryear{Potter, Stadel  \& Teyssier}{Potter
  et~al.}{2017}]{potter2017pkdgrav3}
Potter D.,  Stadel J.,   Teyssier R.,  2017, Computational Astrophysics and
  Cosmology, 4, 1

\bibitem[\protect\citeauthoryear{{Rasia} et~al.,}{{Rasia}
  et~al.}{2015}]{Rasia2015}
{Rasia} E.,  et~al., 2015, \mn@doi [\apjl] {10.1088/2041-8205/813/1/L17}, \href
  {https://ui.adsabs.harvard.edu/abs/2015ApJ...813L..17R} {813, L17}

\bibitem[\protect\citeauthoryear{{Schaye} et~al.,}{{Schaye}
  et~al.}{2015}]{2015MNRAS.446..521S}
{Schaye} J.,  et~al., 2015, \mn@doi [\mnras] {10.1093/mnras/stu2058}, \href
  {https://ui.adsabs.harvard.edu/abs/2015MNRAS.446..521S} {446, 521}

\bibitem[\protect\citeauthoryear{Schmidhuber}{Schmidhuber}{2015}]{schmidhuber2015deep}
Schmidhuber J.,  2015, Neural Networks, 61, 85

\bibitem[\protect\citeauthoryear{{Sembolini}, {Yepes}, {De Petris},
  {Gottl{\"o}ber}, {Lamagna}  \& {Comis}}{{Sembolini}
  et~al.}{2013}]{Sembolini2013}
{Sembolini} F.,  {Yepes} G.,  {De Petris} M.,  {Gottl{\"o}ber} S.,  {Lamagna}
  L.,   {Comis} B.,  2013, \mn@doi [\mnras] {10.1093/mnras/sts339}, \href
  {https://ui.adsabs.harvard.edu/abs/2013MNRAS.429..323S} {429, 323}

\bibitem[\protect\citeauthoryear{{Skillman}, {Warren}, {Turk}, {Wechsler},
  {Holz}  \& {Sutter}}{{Skillman} et~al.}{2014}]{skillman2014darksky}
{Skillman} S.~W.,  {Warren} M.~S.,  {Turk} M.~J.,  {Wechsler} R.~H.,  {Holz}
  D.~E.,   {Sutter} P.~M.,  2014, arXiv e-prints, \href
  {https://ui.adsabs.harvard.edu/abs/2014arXiv1407.2600S} {p. arXiv:1407.2600}

\bibitem[\protect\citeauthoryear{{Stiskalek}, {Bartlett}, {Desmond}  \&
  {Anbajagane}}{{Stiskalek} et~al.}{2022}]{StiskalekScatter2022}
{Stiskalek} R.,  {Bartlett} D.~J.,  {Desmond} H.,   {Anbajagane} D.,  2022,
  \mn@doi [\mnras] {10.1093/mnras/stac1609}, \href
  {https://ui.adsabs.harvard.edu/abs/2022MNRAS.514.4026S} {514, 4026}

\bibitem[\protect\citeauthoryear{{Sunyaev} \& {Zeldovich}}{{Sunyaev} \&
  {Zeldovich}}{1972}]{SZeffect}
{Sunyaev} R.~A.,  {Zeldovich} Y.~B.,  1972, Comments on Astrophysics and Space
  Physics, \href {https://ui.adsabs.harvard.edu/abs/1972CoASP...4..173S} {4,
  173}

\bibitem[\protect\citeauthoryear{{Truong} et~al.,}{{Truong}
  et~al.}{2018}]{Truong2018}
{Truong} N.,  et~al., 2018, \mn@doi [\mnras] {10.1093/mnras/stx2927}, \href
  {https://ui.adsabs.harvard.edu/abs/2018MNRAS.474.4089T} {474, 4089}

\bibitem[\protect\citeauthoryear{Tully \& Fisher}{Tully \&
  Fisher}{1977}]{tully1977new}
Tully R.~B.,  Fisher J.~R.,  1977, Astronomy and Astrophysics, 54, 661

\bibitem[\protect\citeauthoryear{{Villaescusa-Navarro}
  et~al.,}{{Villaescusa-Navarro} et~al.}{2022}]{CAMELS}
{Villaescusa-Navarro} F.,  et~al., 2022, arXiv e-prints, \href
  {https://ui.adsabs.harvard.edu/abs/2022arXiv220101300V} {p. arXiv:2201.01300}

\bibitem[\protect\citeauthoryear{Virtanen et~al.,}{Virtanen
  et~al.}{2020}]{2020SciPy-NMeth}
Virtanen P.,  et~al., 2020, \mn@doi [Nature Methods]
  {10.1038/s41592-019-0686-2}, \href {https://rdcu.be/b08Wh} {17, 261}

\bibitem[\protect\citeauthoryear{{Vogelsberger} et~al.,}{{Vogelsberger}
  et~al.}{2014}]{vogelsberger2014illustris}
{Vogelsberger} M.,  et~al., 2014, \mn@doi [\mnras] {10.1093/mnras/stu1536},
  \href {https://ui.adsabs.harvard.edu/\#abs/2014MNRAS.444.1518V} {444, 1518}

\bibitem[\protect\citeauthoryear{Wadekar, Villaescusa-Navarro, Ho  \&
  Perreault-Levasseur}{Wadekar et~al.}{2021}]{Wadekar_2021neutralhydrogen}
Wadekar D.,  Villaescusa-Navarro F.,  Ho S.,   Perreault-Levasseur L.,  2021,
  \mn@doi [The Astrophysical Journal] {10.3847/1538-4357/ac033a}, 916, 42

\bibitem[\protect\citeauthoryear{{Wadekar} et~al.,}{{Wadekar}
  et~al.}{2022}]{CAMELSSZ}
{Wadekar} D.,  et~al., 2022, arXiv e-prints, \href
  {https://ui.adsabs.harvard.edu/abs/2022arXiv220101305W} {p. arXiv:2201.01305}

\bibitem[\protect\citeauthoryear{Wu, Evrard, Hahn, Martizzi, Teyssier  \&
  Wechsler}{Wu et~al.}{2015}]{wu2015rhapsody}
Wu H.-Y.,  Evrard A.~E.,  Hahn O.,  Martizzi D.,  Teyssier R.,   Wechsler
  R.~H.,  2015, Monthly Notices of the Royal Astronomical Society, 452, 1982

\bibitem[\protect\citeauthoryear{{Zandanel}, {Fornasa}, {Prada}, {Reiprich},
  {Pacaud}  \& {Klypin}}{{Zandanel} et~al.}{2018}]{zandanell2018}
{Zandanel} F.,  {Fornasa} M.,  {Prada} F.,  {Reiprich} T.~H.,  {Pacaud} F.,
  {Klypin} A.,  2018, \mn@doi [\mnras] {10.1093/mnras/sty1901}, \href
  {https://ui.adsabs.harvard.edu/abs/2018MNRAS.480..987Z} {480, 987}

\bibitem[\protect\citeauthoryear{Zhang, Liu, Zhang  \& Almpanidis}{Zhang
  et~al.}{2017}]{comparison2017zhang}
Zhang C.,  Liu C.,  Zhang X.,   Almpanidis G.,  2017, Expert Systems with
  Applications, 82, 128

\makeatother
\end{thebibliography}

\appendix

\section{Description and enumeration of feature variables}
\label{appendixA}

In this appendix, we describe the selected 26  features from the {\sc Rockstar + Consistent Trees} catalogues. Although this information can be found in \cite{Rockstar} and \cite{2013ApJ...763...18B}, as well as  in  the CosmoSim Multidark database \url{https//www.cosmosim.org/}, we include in \autoref{table:features} a brief description of the  variables,  for the reader's convenience. 

\begin{figure*}
\includegraphics[width=1.\textwidth]{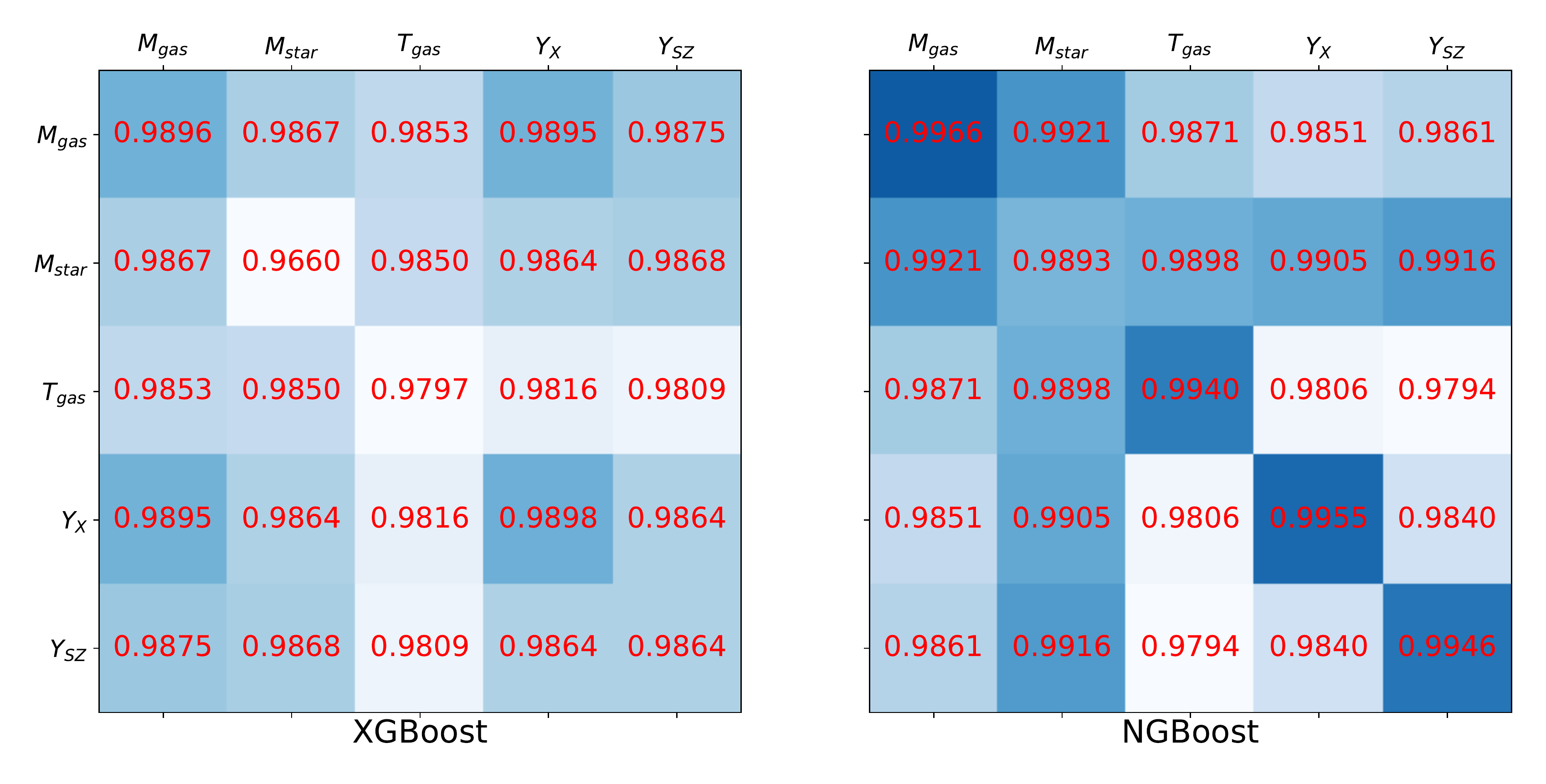}
\caption{The covariance matrices, $\mathcal{Q}$, of  predicted target values   for XGBoost and NGBoost, normalised to the covariance of the ground truth data from The300 simulations. }
\label{fig:cov}
\end{figure*}

\begin{table*}
\caption{The feature variables used in this text from the {\sc Rockstar} catalogue. The first column represents the variable name and their enumeration in brackets.  }\label{table:features}
\begin{tabular}{@{}lll@{}}
\toprule
Variable                       & Units    & Description                                                                                                                 \\ \midrule
M2500c (0)                     & $\hMsun$ & Mass inside a radius of a sphere where the matter density is 2500 times the critical density at the  cluster's redshift \\
num\_prog (1)                  &          & total number of progenitors of the cluster                                                                                  \\
M500c (2)                      & $\hMsun$ & Mass inside a radius of a sphere where the matter density is 500 times the critical density at the  cluster's redshift  \\
M200c (3)                      & $\hMsun$ & Mass inside a radius of a sphere where the matter density is 200 times the critical density at  the cluster's redshift  \\
Mpeak (4)                      & $\hMsun$ & The  peak value of the halo mass across  its accretion history                                                \\
mvir (5)                       & $\hMsun$ & halo mass within the virial radius                                                                                              \\
Macc (6)                       & $\hMsun$ & halo mass at  accretion time.                                                                               \\
Vpeak (7)                      & km/s     & Peak value of  Vmax(9) across mass  accretion history.                                                                                           \\
Vmax\textbackslash{}@Mpeak (8) & km/s     & Vmax at the expansion time at which Mpeak was reach                                                                                  \\
Vmax (9)                       & km/s     & maximum value of the circular velocity.                                                                                                 \\
Vacc (10)                      & km/s     & Vmax at accretion time                                                                                                     \\
rvir (11)                      & \hkpc   & halo radius at virial overdensity  \\
vrms (12)                      & km/s     & root mean squared velocity dispersion                                                                                     \\
b\_to\_a(500c) (13)            &          & ration between the second largest shape ellipsoid axis and largest shape ellipsoid axis, for particles within $R_{500}$      \\
c\_to\_a(500c) (14)            &          & ration between the third largest shape ellipsoid axis and largest shape ellipsoid axis, for particles within $R_{500}$       \\
b\_to\_a (15)                  &          & ration between the second largest shape ellipsoid axis and largest shape ellipsoid axis determined by method in \cite{allgood2006}                                \\
c\_to\_a (16)                  &          & ration between the third largest shape ellipsoid axis and largest shape ellipsoid axis  determined by method in \cite{allgood2006} \\
rs (17)                        & \hkpc     & comoving scale radius  from the    fit to a  NFW \citep{1997ApJ...490..493N} density profile                                                                                                     \\
Rs\_Klypin (18)                & \hkpc     & comoving scale radius determined using Vmax and Mvir formula  \citep{2011ApJ...740..102K}                                                                        \\
T/|U| (19)                     &          & the ratio between the total  kinetic and potential energies of particles within virial radius.                                                                          \\
Xoff (20)                      & \hkpc  & Offset between  comoving density peak  and the particles center of mass  position                                                                    \\
Voff (21)                      & km/s     &  Offset between halo core velocity and the  center of mass velocity for particles within the virial radius \\
Spin (22)                      &          & Peebles's dimensionless Spin parameter of the halo \citep{1969ApJ...155..393P}.                                                                                                  \\
Spin\_Bullock (23)             &          & Bullock's dimensionless  spin parameter \citep{bullockspin}                                                  \\
a (24)                         &          & Expansion scale factor of the corresponding simulation snapshot                                                                       \\
scale\_of\_last\_MM (25)       &          & Expansion scale factor of the last major merger with a mass ratio greater than 0.3                                                      \\
Halfmass\_Scale (26)           &          & Expansion scale factor when the most massive halo progenitor reached $0.5\times \text{Mpeak(4)}$                                                       \\ \bottomrule
\end{tabular}
\end{table*}

\section{Covariance structure of baryonic targets}
\label{appendixB}

In \autoref{sec:NGBvsXGB} we have estimated the scatter for different targets binning in halo mass $M_{500}$ intervals. In this appendix we check that the whole covariance structure of the ML-predicted baryonic properties is similar to that of The300 simulation. In order to do that, we have computed the covariance matrix as
\begin{equation}
    \text{Cov}(y_i,y_j)= E[(y_i-E(y_i))(y_j-E(y_j))]
\end{equation}
and the element-wise quotient between the covariance matrix of ML predicted targets and the true targets, i.e. :
\begin{equation}
    \mathcal{Q}=\frac{\text{Cov}_{\text{pred}}}{\text{Cov}_{\text{true}}}=\frac{\text{Cov}(y_{i,\text{pred}},y_{j,\text{pred}})}{\text{Cov}(y_{i,\text{true}},y_{j,\text{true}})}.
\end{equation}
In this way, $\mathcal{Q}$ shows how similar the covariance matrix of the predicted targets is with respect to the corresponding one of true targets. $\mathcal{Q}$ is plotted in \autoref{fig:cov} for both models XGBoost and NGBoost. As can bee seen in the figure, $\mathcal{Q}$ values are very close to 1 but always less than 1.  This means that the predictions of both ML models  are distributed around the  mean  true values, but they do not completely reproduce  the  tails of the distributions of the real data. As a general result, NGBoost baryonic properties show  a  covariance structure  closer to the corresponding  ground-truth values of The300 simulations.



\bsp	
\label{lastpage}

\end{document}